\documentclass[a4paper,onecolumn,11pt,accepted=2021-11-23]{quantumarticle}
\pdfoutput=1
\usepackage[utf8]{inputenc}
\usepackage[english]{babel}
\usepackage[T1]{fontenc}
\usepackage{amsmath}
\usepackage{hyperref}
\usepackage{float}

\usepackage{tikz}
\usepackage{lipsum}

\hypersetup{colorlinks = true, pdftitle = SSMC}

\usepackage{graphicx}
\usepackage{dcolumn}
\usepackage{bm}
\usepackage{mathtools}          
\usepackage{braket}
\usepackage{color}
\usepackage[draft]{changes}
\newcommand{\norm}[1]{\left\lVert #1 \right\rVert}
\newcommand{\cond}{{\rm cond} }

\usepackage[numbers,sort&compress]{natbib}

\begin{document}

\title{Sequential optical response suppression for chemical mixture characterization}

\author{Alicia B. Magann}
\email{abmagan@sandia.gov}

\affiliation{Department of Chemical \& Biological Engineering, Princeton University, Princeton, New Jersey 08544, USA}

\affiliation{Center for Computing Research, Sandia National Laboratories, Albuquerque, New Mexico 87185, USA}

\author{Gerard McCaul}
\email{gmccaul@tulane.edu}
\affiliation{Department of Physics, Tulane University, New Orleans, LA 70118, USA}

\author{Herschel A. Rabitz}
\email{hrabitz@princeton.edu}
\affiliation{Department of Chemistry, Princeton University, Princeton, New Jersey 08544, USA}

\author{Denys I. Bondar}
\email{dbondar@tulane.edu}
\affiliation{Department of Physics, Tulane University, New Orleans, LA 70118, USA}

\begin{abstract}
The characterization of mixtures of non-interacting, spectroscopically similar quantum components has important applications in chemistry, biology, and materials science. We introduce an approach based on quantum tracking control that allows for determining the relative concentrations of constituents in a quantum mixture, using a single pulse which enhances the distinguishability of components of the mixture and has a length that scales linearly with the number of mixture constituents. To illustrate the method, we consider two very distinct model systems: mixtures of diatomic molecules in the gas phase, as well as solid-state materials composed of a mixture of components. A set of numerical analyses are presented, showing strong performance in both settings. 
\end{abstract}

\maketitle
\section{\label{sec:level1} Introduction}

Across the sciences, methods for determining the concentrations of a mixture's components are in high demand. The characterization of such mixtures can be a challenging task, especially if the components have very similar chemical and physical properties. This can occur for example in mixtures of isomers, isotopes, or chiral molecules, as well as in mixtures of larger biological molecules such as amino acids or proteins. For example, spectral crosstalk currently limits the ability to distinguish between different fluorescent proteins in experiments \cite{lichtman2005fluorescence}, but if this challenge could be overcome such that the accurate characterization of fluorescent protein mixtures were possible, this would constitute a significant advance with implications across the biological sciences, including in synthetic biology \cite{medley2005simultaneous}, neuroscience \cite{feng2000imaging}, and cytometry \cite{bagwell1993fluorescence,speicher1996karyotyping,perfetto2004seventeen}. 

At present, standard methods (such as chromatography \cite{wilson2000encyclopedia}) often struggle to discriminate accurately between different components, as they typically rely on the ability to resolve specific differences (e.g. in absorption peaks) between the species, which may not be possible for mixtures composed of very similar constituents. Optimal dynamic discrimination (ODD) was developed to address this challenge \cite{oddrabitz,li2005optimal}, by leveraging ``photonic reagents'' in the form of optimally-shaped fields, and has been demonstrated in proof-of-principle experiments in several systems, including dye molecules \cite{brixner2001photoselective}, flavin molecules \cite{PhysRevLett.102.253001,roslund2011resolution}, amino acids \cite{rondi2012coherent}, and fluorescent proteins \cite{2016NatSR...625827G}. In essence, ODD makes use of the fact that as long as distinctions are present between the Hamiltonians of the different components, they will evolve differently under the influence of a driving field applied to the mixture. These differences in the dynamics and the resulting optical responses can be learned, and the premise of ODD is to optimize a field to amplify these differences over time, thereby allowing for discrimination between the components. The procedure for accomplishing this is based on quantum optimal control, and proceeds via an iterative optimization loop in the laboratory. 

In this article, we introduce a new procedure for characterizing mixtures of non-interacting components based on quantum tracking control principles, which we term \emph{scalable, single-pulse mixture characterization} (SSMC). Quantum tracking control is a method for designing a field $E(t)$, $t\in[0,t_f]$, to drive the evolution of an observable expectation value along a prescribed time-dependent trajectory or ``track'' over a time interval $t\in[0,t_f]$. If reliable models for similar systems are available, the field $E(t)$ can be computed numerically by solving a modified dynamical equation at each time step, such that the full field can be determined from a single forward evolution of the system dynamics. This tracking control approach stands in contrast to numerical field-design procedures based on quantum optimal control that identify fields for achieving a control objective in a system via an expensive iterative optimization procedure, where at each iteration a forward evolution of the system dynamics is required, and hundreds or thousands of iterations can be needed for convergence of the objective function. Numerical studies of tracking control have been performed for numerous systems including a qubit \cite{Lidar2004}, atomic systems \cite{PhysRevLett.118.083201}, diatomic and triatomic molecules \cite{gross1993inverse,Chen1995,Chen1997}, molecular rotors \cite{PhysRevA.98.043429}, and many-body solid-state systems \cite{mccaul2020driven,mccaul2020controlling}.

In general, quantum tracking control can suffer from singularities appearing in $E(t)$ due to the noninvertibility of the associated dynamical equation. Techniques to manage singularities in this context were introduced in \cite{doi:10.1063/1.477857,doi:10.1063/1.1582847}, and more recently, it was found that singularities could be avoided altogether for certain tracking control problems \cite{PhysRevLett.118.083201,PhysRevA.98.043429,mccaul2020driven,mccaul2020controlling}. In these latter cases, quantum tracking control offers a computationally attractive method for designing fields to control observable dynamics. 

In SSMC, the task of characterizing mixtures is formulated as one such singularity-free tracking problem, where the controlled observable is taken to be the optical response, and the prescribed track is defined such that the optical responses of all components in the mixture are sequentially turned off, effectively making each species sequentially invisible for a specified time interval. SSMC is a strategy directed towards systems where either explicit models are available, or effective models can be estimated from experimental data. In practice, such models will always contain some degree of error, and in principle the SSMC fields could be used as a starting basis for ODD iterative enhancement in the laboratory. To this end, we present a general model-based procedure for deriving equations for fields that drive the optical response along such tracks. When these fields are applied, the time-dependent optical response data from the full mixture can be measured and used to evaluate the relative concentrations of each of the components by solving a linear system of equations.

The primary benefit of the SSMC procedure is its simplicity. Namely, the use of tracking control enables the field amplitudes to be determined by solving a simple analytical expression at each time step. And, while the SSMC procedure could be implemented using other forms of quantum control (i.e. optimal control) \cite{Schroder2008,Mundt_2009}, this would incur significant additional computational costs, and should only be considered when tracking control cannot be made singularity-free. Another key feature of this procedure is its scalability with respect to the number of mixture constituents. That is, the field used to accomplish the characterization has a length that scales linearly in the number of components of the mixture under consideration, and the computational effort required $\emph{a priori}$ and $\emph{post facto}$ to determine the relative concentrations is also efficient with respect to the number of components (see Fig. \ref{fig:SSMCscaling}, below).

To highlight the generality of SSMC, we present proof-of-principle numerical illustrations involving two radically different models: mixtures of diatomic molecules in the gas phase, as well as solid-state materials composed of a mixture of components. The SSMC procedure is shown to perform consistently well in both settings. As a comparison we consider the use of another single-pulse strategy, namely the use of transform-limited pulses, and show that SSMC pulses yield superior performance in all cases considered.

The remainder of this article is organized as follows. To begin, in Section \ref{Sec:Theory}, the general SSMC procedure is introduced. Then, in Section \ref{sec:molcharacterisation}, the application of SSMC to the characterization of mixtures of diatomics is considered, and a series of numerical analyses are presented. To demonstrate the flexibility of SSMC, this is followed in Section \ref{Sec:SolidState} by an application of SSMC to solid-state systems, where associated numerical illustrations are provided. We then conclude in Section \ref{Sec:Conclusion} with further discussion and outlook.

\section{Theory} \label{Sec:Theory}

Consider the optical response of a mixture of $n_s$ non-interacting quantum components, whose individual optical responses at some time $t$ are denoted $R^{(s)}(t)$, for $s = 1,\cdots,n_s$. The optical response of the mixture is simply given by a weighted sum over all species' optical responses:

\begin{equation}
R_{\text{mix}}(t)=\sum_{s=1}^{n_s}y_sR^{(s)}(t)\added{,}
\label{Eq:Rmix}
\end{equation}
where $y_s$ denotes the relative concentration of component $s$. 

\begin{figure}[t]
        \centering
        \includegraphics[width=0.7\columnwidth]{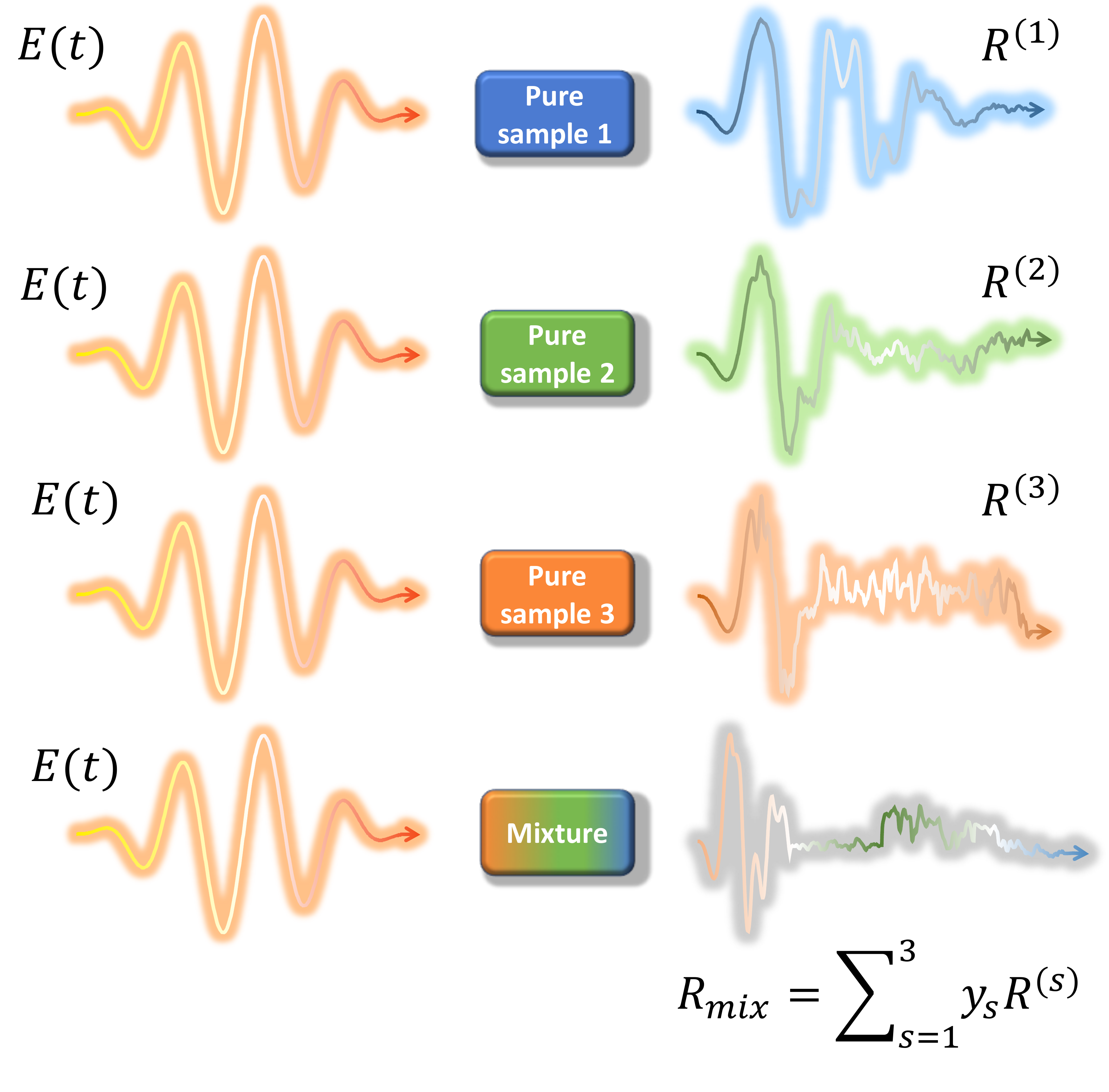}
        \caption{Diagram of naive characterization procedure using a transform-limited pulse with a simple form. The diagram shows an input pulse $E(t)$, which serves as a ``photonic reagent'' for inducing optical responses $R^{(1)}$, $R^{(2)}$, and $R^{(3)}$ in species 1, 2, and 3, respectively. The optical response of a mixture of these species, $R_{\text{mix}}$, is given by the weighted sum of the species optical responses, where the weights are given by the relative concentrations, $y_1$, $y_2$, and $y_3$, of the species. }
        \label{Fig:NaiveProcedure}
\end{figure}

A naive characterization procedure might involve the application of a transform-limited pulse to the mixture (see Fig. \ref{Fig:NaiveProcedure}), in order to induce an optical response from the mixture as per Eq. (\ref{Eq:Rmix}), which could be measured and recorded. Then, if each species optical response $R^{(s)}$ were known \emph{a priori}, the measured optical response data from the mixture could be utilized to determine the relative species concentrations by inverting Eq. (\ref{Eq:Rmix}) with respect to the vector $(y_1,\cdots,y_{n_s})^T$, i.e., by solving the set of coupled equations

\begin{equation} 
   \mathbf{A} \mathbf{\bar{y}} = \mathbf{R_{mix}} \added{,}
    \label{lstsqs}
\end{equation}
where $\mathbf{A} = (\mathbf{R}^{(1)},\,\mathbf{R}^{(2)}, \cdots,\mathbf{R}^{(n_s)})$ is a rectangular matrix containing $\mathbf{R}^{(s)}$, $s=1,\cdots,n_s$, which denote the optical responses of each of the individual species; $\mathbf{\bar{y}} = (\bar{y}_1,\, \bar{y}_2, \cdots, \bar{y}_{n_s})^T$ is a vector containing the estimated values of the relative species concentrations, where the bars are used to distinguish between the estimated values $\bar{y}_s$ and the true values $y_s$; and $\mathbf{R_{mix}}$ is a vector containing the mixture optical response values. The boldface vector notation is utilized in reference to solving the system of equations associated with the mixture characterization problem (\ref{lstsqs}), in order to denote data with time-dependent entries. When describing the underlying physics in the material below, we do not use boldface notation.

Eq. (\ref{lstsqs}) is overdetermined, but may still be solved by (for example) using a least squares approach. The effectiveness of inverting $\mathbf{A}$ can be quantified using the condition number, 
\begin{equation}
    \textrm{cond}(\mathbf{A}) = \frac{\sigma_{max}(\mathbf{A})}{\sigma_{min}(\mathbf{A})}.
    \label{Eq:CondA}
\end{equation}
Here, $\sigma_{max}(\mathbf{A})$ and $\sigma_{min}(\mathbf{A})$ denote the largest and smallest singular values of $\mathbf{A}$. The larger the condition number, then the more ill-conditioned the matrix $\mathbf{A}$ is, meaning numerical solutions of Eq. (\ref{lstsqs}) will be less accurate for estimating the relative concentrations. In order to quantify the error $\epsilon$ associated with the estimate of the relative concentrations, the 2-norm can be used as per
\begin{equation}
    \epsilon = \norm{\mathbf{y}-\bar{\mathbf{y}}}_2\added{.}
    \label{Eq:Epsilon}
\end{equation}
If the constituents of a mixture possess similar chemical, physical, and optical properties, one can expect that their optical responses collected in $\mathbf{A} $ will also be similar, leading to large condition numbers $\textrm{cond}(\mathbf{A}) $ that result in significant errors $\epsilon$ in the estimated relative concentrations found by solving Eq. (\ref{lstsqs}). 

The premise of SSMC is to design individual optical responses such that the distinguishability between the optical responses of different species collected in $\mathbf{A} $ is enhanced, thereby yielding a better-conditioned system of equations in Eq. (\ref{lstsqs}) that can be solved to characterize the mixture with better accuracy. In the remainder of this section, we present the SSMC procedure as a method for achieving more accurate and robust characterization.

The SSMC procedure begins by applying a pump pulse to the mixture, denoted by $E_p(t)$, in order to induce an initial response in the mixture. We note that as the only role of the pump pulse is to induce a response, there is significant freedom to choose it such that other desired properties can be imposed upon the mixture (e.g., alignment of the mixture components). In order to promote \emph{distinguishability} between optical responses of the different species, the optical responses of each of the $n_s$ components are then turned off in series for some time period $T$ through the sequential application of $n_s$ pulses $E_1(t),\cdots,E_{n_s}(t)$. The pulse $E_s(t)$ is designed to suppress the optical response of a particular species $s$ such that $R^{(s)}(t) = 0$ for the duration of the pulse $E_s(t)$. 

A first step in determining the $E_1(t),\cdots,E_{n_s}(t)$ that will suppress $R^{(1)}(t),\cdots R^{(n_s)}(t)$ is the derivation of an equation describing the form of each $R^{(s)}(t)$ in terms of $E_s(t)$ and one or multiple observable expectation values at time $t$. The derivation of such equations relies on models for the mixture components; for example molecular systems and solid-state systems, see Sections \ref{sec:molcharacterisation} and \ref{Sec:SolidState}, respectively, while a more general derivation is given in Appendix \ref{Appendix}. 

Once it has been derived, the equation for $R^{(s)}\big(E_s(t),\cdots\big) = 0$ can simply be inverted to produce an expression for $E_s(t)$. This inversion procedure is the premise of quantum tracking control, which underlies SSMC. It is important to note that in general, the response will be a highly nonlinear function of the driving field. As such it is difficult to obtain an interpretation to the action of the field $E_s(t)$ (however, formal mechanism analyses for SSMC applications could be done in principle, using techniques from e.g. \cite{mitra2003identifying,mitra2004mechanistic,mitra2006quantum}).

The full SSMC field $E(t)$ is then formed by concatenating the pump pulse $E_p(t)$, with each of these $n_s$ pulses, yielding one pulse 
\[
  E(t) =
  \begin{cases}
 E_p(t) & \text{if $t\in[0,T)$} \\
E_1(t) & \text{if $t\in[T,2T)$} \\
E_2(t) & \text{if $t\in[2T,3T)$} \\
 \cdots & \\
  E_{n_s}(t)& \text{if $t\in[n_sT,(n_s+1)T)$}
  \end{cases}
\]
that achieves the goal of suppressing the optical responses of each mixture constituent in series, thereby enhancing the distinguishability of the elements in $\mathbf{A} $.

The design of the SSMC field $E(t)$ need only be done once for a given combination of species, and importantly, the resultant SSMC pulse $E(t)$ is completely independent of the values of $y_s$, $s=1,\cdots n_s$ of the species in the mixture. Hence, regardless of the relative species concentrations in a given mixture, the same pulse can always be used to determine these concentrations. 

\begin{figure} 
    \centering
        \includegraphics[width=0.9\columnwidth]{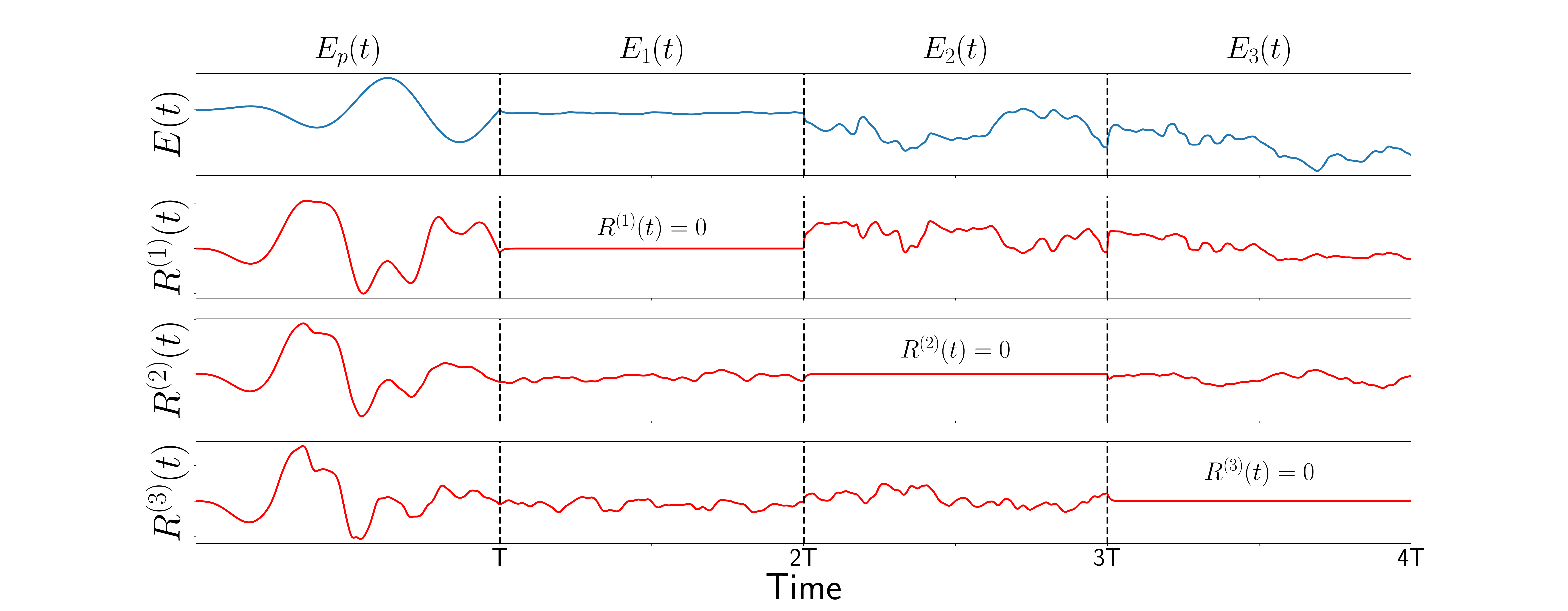}
        \caption{Depiction of SSMC procedure. Following the application of the pump pulse $E_p(t)$, the pulses $E_1(t),\cdots,E_{n_s}(t)$ are applied to sequentially switch off the optical response of each species for a time period $T$. The resultant library pulse $E(t)$, formed by a concatenation of these pulses, can then be used on an unknown mixture to determine the relative concentrations of the constituents.  \label{fig:3speciesexample}
}
\end{figure}

After they are formed in this manner, the SSMC field $E(t)$, along with the associated optical responses of the components $R^{(s)}(t)$, $s=1,\cdots,n_s$, $t\in (0, (n_s+1)T]$, can be stored in a library, where a sample entry involving three components might look like Fig. \ref{fig:3speciesexample}.\footnote{We remark that the sample library pulse shown was formed by a simple concatenation, and that this can lead to sharp variations in the field at the concatenation points. A variety of approaches could be employed to improve the smoothness of the fields at these points, such as adding a short intervening period between segments and using this period to enforce a smooth connection between them.}

In the laboratory, $E(t)$ serves as a photonic reagent that can be drawn from the library, in analogy with how a chemical reagent might be drawn from a stockroom, and applied to a mixture to determine the relative concentrations of the different components using a single pulse. Over the course of the pulse, the mixture's time-dependent optical response data can be recorded via measurements and stored as a set of discrete entries in the vector $\mathbf{R_{mix}}$. For each of the $n_s$ pulses $E_1(t),\cdots E_{n_s}(t)$, we denote the number of entries in the associated optical response vectors by $n_t$, such that the total number of entries for each species over the course of the SSMC procedure is $(1+n_s)n_t$. In order to determine the concentrations, the data associated with the pump pulse is discarded, and only $n_s n_t$ entries are utilized. Ultimately, the estimated species concentrations $\bar{y}_s$, $s=1,\cdots,n_s$ can be determined by solving the set of coupled equations in Eq. (\ref{lstsqs}), where in SSMC,
\begin{equation*}
\mathbf{A} = 
\begin{pmatrix}
0 & \mathbf{R}_1^{(2)} & \cdots & \mathbf{R}_1^{(n_s)} \\
\mathbf{R}_2^{(1)} & 0 & \cdots & \mathbf{R}_2^{(n_s)} \\
\vdots  & \vdots  & \ddots & \vdots  \\
\mathbf{R}_{n_s}^{(1)} & \mathbf{R}_{n_s}^{(2)} & \cdots & 0 
\end{pmatrix}
\end{equation*}
is a rectangular $(n_s n_t\times n_s )$ dimensional matrix, where $\mathbf{R}_{1}^{(s)}$ denotes the $(n_t\times 1)$ vector whose elements are the optical response values associated with species $s$ collected for $t\in(T,2T]$ while $E_1(t)$ is applied, $\mathbf{R}_{2}^{(s)}$ is the $(n_t\times 1)$ vector of optical response values for species $s$ collected for $t\in(2T,3T]$, etc, and
\begin{equation}
    \mathbf{R_{mix}} = (\mathbf{R_{mix,1}},\, \mathbf{R_{mix,2}}, \cdots ,\mathbf{R_{mix,n_s}})^T
\end{equation}
is an $(n_s n_t\times 1)$ dimensional vector, where $\mathbf{R_{mix,1}}$ denotes the  $(n_t\times 1)$ vector containing the mixture optical response values collected for $t\in(T,2T]$ while $E_1(t)$ is applied, $\mathbf{R_{mix,2}}$ is the  $(n_t\times 1)$ vector containing mixture optical response values collected for $t\in(2T,3T]$, etc.

Because the SSMC procedure guarantees that the optical response of each species is zero for some period $T$, the matrix $\mathbf{A}$ will be structured to contain zeros, which can be expected to lead to a lower condition number and smaller error. This is in contrast to a more naive approach, where a simple transform-limited pulse is applied to induce an optical response, and the recorded values used to solve a similar set of equations. 

\begin{figure} 
    \centering
        \includegraphics[width=0.6\columnwidth]{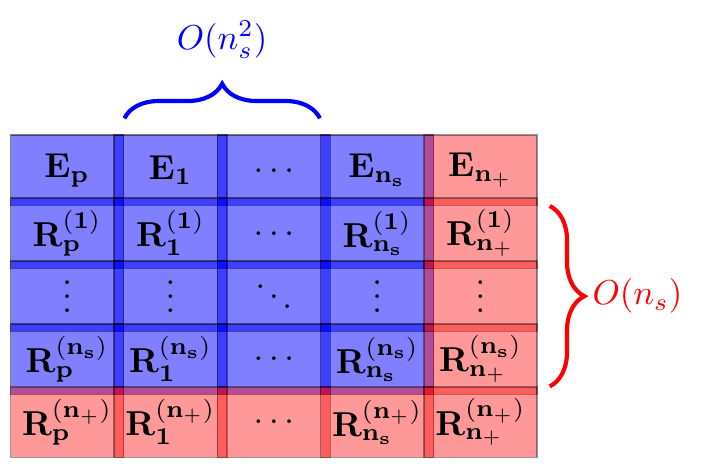}
        \caption{Schematic illustration of the scaling of SSMC (where $n_+=n_{s}+1$). Each element in this block matrix is a vector of length $n_t$ representing a time period $T$. To calculate the discriminating pulse and library of responses for $n_s$ species will have a cost of {\color{blue} $O(n^2_s)$}. Adding a new species to an existing library however only adds a computational cost of {\color{red} $O(n_s)$}. }\label{fig:SSMCscaling}
\end{figure}

An important feature of SSMC is its scalability with respect to the number of mixture constituents. That is, if one wishes to calculate the SSMC field for $n_s$ species, one must calculate a library of $n_s$ different responses of length $(n_s+1)T$, which yields an SSMC field $E(t)$ with a length that scales as $O(n_s)$. Meanwhile, the computational cost for constructing the SSMC pulse and the associated response library is $O(n_s^2)$. If one wishes to extend this library by adding new species however, it is not necessary to completely recalculate the SSMC pulse and species responses. Instead, one simply appends the new species to the library, evolving it with the already calculated pulse (at cost $n_sT$) before applying tracking control to eliminate its response for a period. This length $T$ extension to the SSMC pulse is then applied to each of the other species in the library (provided the final state of each species has been saved) to generate the necessary addition to the response of each species in the library. Adding new species to the discriminating pulse and response library in this manner thus carries a cost of only $O(n_s)$. Both of these scalings are illustrated schematically in Fig.~\ref{fig:SSMCscaling}.  This scalability is of particular value in mixture characterisation, as while the condition number (and therefore error) can scale exponentially with $n_s$ (see Fig. \ref{Fig:MolDiscFig}), the SSMC method for reducing these errors scales only polynomially. Additionally, if one were simply to use a single transform-limited pulse for discrimination rather than SSMC, the scaling would depend on whether the length of the pulse was fixed, or lengthened by $T$. In the case of a fixed length, adding another species would also be of cost $O(n_s)$. This would however \emph{increase} the degree of overdetermination in the system, and therefore further reduce the accuracy of calculated concentrations. If one lengthens the overall time to counteract this however, the transform-limited condition necessitates a modification of the pulse. In this case, the response of each species would need to be recalculated, leading to a cost of $O(n_s^2)$.   

For the purposes of numerical and conceptual simplicity, in what follows we will consider only one dimensional systems. Some essential information (such as orientation) present in a three dimensional system is inevitably lost in these lower dimensional models, but one would expect that random orientations in a mixture would average out these contributions to the optical response of individual molecules. Finally, we note that in principle the SSMC procedure can be generalised to interacting systems, but at the cost of tracing out interacting degrees of freedom when tracking each species. Given the considerable additional analytic and computational complexity necessary to account for interactions, we consider only non-interacting mixtures in this work.

\section{Characterization of molecular mixtures \label{sec:molcharacterisation}}

In this section, we explore an application of the SSMC procedure towards the characterization of mixtures of $n_s$ non-interacting molecular species, as could occur in e.g. atmospheric or combustion samples, where the set of possible molecular constituents in the sample is known, whereas the relative amounts of each are not. In this setting, we consider each species $s$ itself as being composed of $N^{(s)}$ interacting nuclear (i.e. rotational or vibrational), polar degrees of freedom modeled in the Born-Oppenheimer approximation, and here assumed to be aligned along the polarization direction of the applied field $E(t)$\footnote{In practice, the alignment of the mixture components along the same axis as the applied field could be accomplished in a preliminary step (e.g., using the pump pulse), using techniques such those described in \cite{RevModPhys.75.543}.}. The dipole moment of a particular degree of freedom $j$ belonging to a molecular species $s$ is denoted $\mu_j^{(s)}$, and is a function of the vibrational or rotational coordinate $r_j^{(s)}$ associated with this degree of freedom. A typical Hamiltonian governing a single species subsystem is given in the dipole approximation by
\begin{equation}
\begin{aligned}
H^{(s)}(t) = \sum_{j=1}^{N^{(s)}}-\alpha_j^{(s)} \frac{\partial^2}{\partial r_j^{{(s)}^2}}+V^{(s)}-\sum_{j=1}^{N^{(s)}}\mu_j^{(s)}E(t) \added{,}
\label{eq:MolHamiltonian}
\end{aligned}
\end{equation}
for some scalars and $\alpha_1^{(s)},\cdots,\alpha_{N^{(s)}}^{(s)}$ and potential $V^{(s)}=V(r_1^{(s)},\cdots,r_{N^{(s)}}^{(s)})$. Because we assume that both $V^{(s)}$ and $\mu_j^{(s)}$ are functions of the coordinate(s), we assume that $[V^{(s)},\mu_j^{(s)}]=0$ for all $j$. 

The optical response of species $s$ is proportional to the acceleration of its total dipole moment, given here as the sum over the dipole moments corresponding to the different degrees of freedom, i.e., 
\begin{equation}
R^{(s)}=\sum_{k=1}^{N^{(s)}} R_k^{(s)}  = \sum_{k=1}^{N^{(s)}}\frac{d^2\langle \mu_k^{(s)}\rangle_t}{dt^2},
\end{equation}
where expressions for each $R_k^{(s)}$ may be obtained via the Heisenberg equation, yielding the result,
\begin{equation}
\begin{aligned}
R_k^{(s)}&=\frac{d^2\langle \mu_k^{(s)}\rangle_t}{dt^2} =-\frac{\alpha_k^{(s)}}{\hbar^2}\big\langle \mathcal{B}^{(s)}\big\rangle_t +\frac{2\alpha_k^{(s)}}{\hbar^2}\bigg\langle \frac{\partial \mu_k^{(s)}}{\partial r_k^{(s)}}^2\bigg\rangle_t E(t)\added{,}
\end{aligned}
\end{equation}
where 
\begin{equation}
\begin{aligned}
   \mathcal{B}^{(s)} &\equiv  \alpha_k^{(s)}\bigg(\frac{\partial^4\mu_k^{(s)}}{\partial r_k^{(s)^4}}+4\frac{\partial^3\mu_k^{(s)}}{\partial r_k^{(s)^3}}\frac{\partial}{\partial r_k^{(s)}}+4\frac{\partial^2\mu_k^{(s)}}{\partial r_k^{(s)^2}}\frac{\partial^2}{\partial r_k^{(s)^2}}\bigg)+2\frac{\partial \mu_k^{(s)}}{\partial r_k^{(s)}}\frac{\partial V^{(s)}}{\partial r_k^{(s)}}\added{,}  
   \end{aligned}
\end{equation}
using $\langle \cdot\rangle_t \equiv \langle\psi(t)|\cdot|\psi(t)\rangle$. The optical response of the full species $s$ containing $N$ degrees of freedom is then given by the sum of the optical responses over all degrees of freedom, as per
\begin{equation}
R^{(s)}=\sum_{k=1}^{N^{(s)}} R_k^{(s)} =\sum_{k=1}^{N^{(s)}}\frac{\alpha_k^{(s)}}{\hbar^2}\bigg\{ -\big\langle \mathcal{B}^{(s)} \big\rangle_t +2\Bigg\langle \frac{\partial \mu_k^{(s)}}{\partial r_k^{(s)}}^2\Bigg\rangle_t E(t)\bigg\}\added{.}
\label{Rs}
\end{equation}
In order to obtain an expression for the laser pulse $E_s(t)$ that suppresses the optical response of species $s$, we may invert the equation $R^{(s)}(t)=0$, with $R^{(s)}$ given in Eq. (\ref{Rs}), to yield the pulse\footnote{The denominator of Eq. (\ref{eq:molfield}) cannot change sign as it contains a sum of $N^{(s)}$ non-negative expectation values. And, the likelihood of encountering a singularity, i.e., where all of these expectation values are simultaneously equal to zero, is very small and never occurred in our numerical analyses. The fact that this possibility exists at all is an artifact of working with a model in the Born-Oppenheimer approximation, and the more general and singularity-free case is treated in the appendix.}
\begin{equation} 
E_s(t)=\frac{\sum_{k=1}^N\frac{\alpha_k^{(s)}}{\hbar^2}\big\langle \mathcal{B}^{(s)} \big\rangle_t }{2 \Big\{ \sum_{k=1}^N\frac{\alpha_k^{(s)}}{\hbar^2}\Big\langle \frac{\partial \mu_k^{(s)}}{\partial r_k^{(s)}}^2\Big\rangle_t\Big\} } \added{.}
\label{eq:molfield}
\end{equation}
Then the resulting dynamical equation governing the mixture,
\begin{equation}
    \frac{d}{dt}|\psi(t)\rangle = -i (H_0 - \mu E_s(\psi,t))|\psi(t)\rangle,
    \label{Eq:NonlinTDSE}
\end{equation}
is highly nonlinear due to the functional dependence of the field $E_s(t)$ on the state $|\psi(t)\rangle$. As such, in practice, the coupled Eqs. (\ref{eq:molfield}) and (\ref{Eq:NonlinTDSE}) can be solved as follows. After evolving $|\psi(t)\rangle$ to time $t=T$ under the influence of a pump pulse $E_p(t)$, begin with an initial value for $E_1(t=T)$ found by evaluating Eq. (\ref{eq:molfield}) with the initial condition $|\psi(T)\rangle$. The state of the mixture is then propagated forward in time via Eq. (\ref{Eq:NonlinTDSE}) by evolving the state of each species independently under its respective Hamiltonian forward a single time step as $|\psi_s(T)\rangle \rightarrow |\psi_s(T+\Delta t)\rangle$ for some species $s$. The updated states are then substituted back into Eq. (\ref{eq:molfield}) to determine $E_1(t=T+\Delta t)$, and so forth, up until time $t=2T$. This procedure is then repeated to determine each of the subsequent fields $E_2(t),\cdots,E_{n_s}(t)$.

\subsection{Numerical illustrations}\label{Sec:molecularnumerics}

\begin{figure}
\centering
\includegraphics[width=0.6\columnwidth]{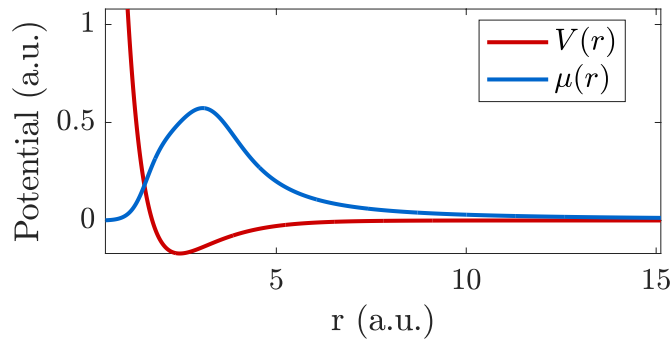}
\caption{Anharmonic Morse potentials $V(r)$ (red curve) and dipole potential $\mu(r)$ (blue curve) as a function of the bond coordinate $r$.}
\label{Fig:Potentials}
\end{figure}

\begin{figure}[t]
     \centering
        \includegraphics[width=1\linewidth]{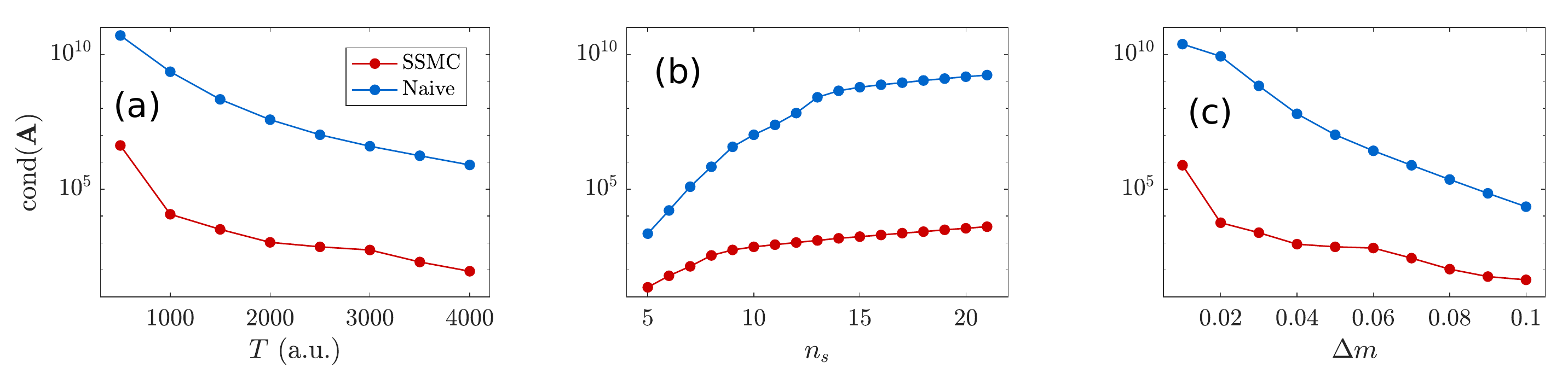}
        \caption{The dependence of $\text{cond}(\mathbf{A})$ on different quantities it shown on semilog plots for the SSMC approach (red) and a naive approach (blue). In all cases, the SSMC procedure leads to better conditioned problems compared with using the naive approach, as indicated by values of $\textrm{cond}(\mathbf{A})$ that are consistently a few orders of magnitude lower for the SSMC procedure compared with the naive procedure. In (a), the dependence on the duration of each pulse $T$ is plotted, taking $n_s = 10$ and $\Delta m=0.05$. In (b), the dependence on the number of species $n_s$ is plotted for $T = 2,500$ a.u. and $\Delta m = 0.05$. In (c), the dependence on the relative mass difference $\Delta m$ is plotted for $T = 2,500$ a.u. and $n_s = 10$.}
        \label{Fig:MolDiscFig}
\end{figure}

\begin{figure}[t]
    \centering
        \includegraphics[width=0.6\columnwidth]{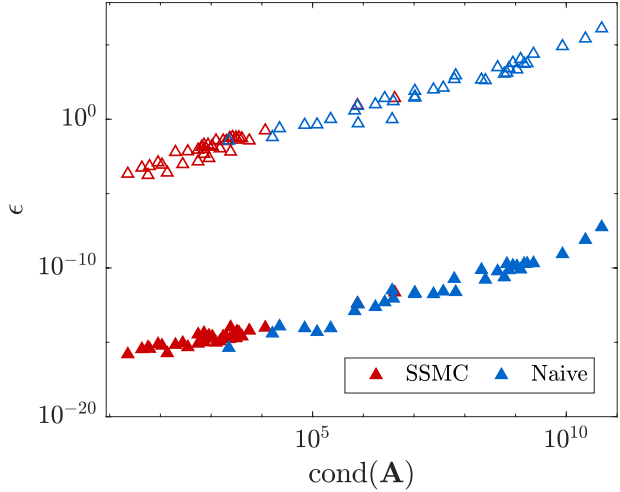}
    \caption{The relation between the condition of the problem $\textrm{cond}(\mathbf{A})$ and the error in the solution $\epsilon$ is shown. Filled point markers correspond to the aggregated data from Fig. \ref{Fig:MolDiscFig}(a)-(c), while unfilled point markers correspond to the same data when $\mathbf{R_{mix}}$ is infected by noise, leading to higher values of $\epsilon$. A clear, approximately linear correlation between $\textrm{cond}(\mathbf{A})$ and $\epsilon$ is evident in both cases.  \label{fig:EpsVsCond}}
\end{figure}

As an illustration, we consider characterizing mixtures of gas-phase diatomic molecules which differ only in their masses, as occurs when discriminating between molecules containing different isotopes. This example is meant to serve as a simple, proof-of-concept application study of SSMC towards molecular systems. We model the vibrational dynamics of each of the diatomics as nonrotating Morse oscillators, initially in their ground vibrational state and aligned with the polarization direction of the applied field \cite{PhysRev.34.57}, with the Hamiltonian for each species given by Eq. (\ref{eq:MolHamiltonian}) with $N^{(s)}=1$ and $\alpha^{(s)} = \frac{\hbar^2}{2m^{(s)}}$ where the reduced mass of $m^{(s)}$ is taken to be a variable parameter \footnote{The vibrational dynamics are simulated in coordinate space using a 1D grid with 100 points, evenly spaced between $r=0.25$ and $r = 12.25$ a.u., with differential operators represented using finite difference formulas and with $\Delta t = 2.5$ a.u. Changes to the grid did not result in qualitative changes to the results presented in Figs. \ref{Fig:MolDiscFig} and \ref{fig:EpsVsCond}, or to the discussions and conclusions in Sec. \ref{sec:molcharacterisation}. Codes available upon reasonable request to the authors. }. In particular, we consider mixtures of components whose masses $m^{(1)},\cdots,m^{(n_s)}$ take on values evenly spaced between $(1-\Delta m)m_{ref}$ and $m_{ref}$, where we choose $m_{ref} = 1800$ a.u.\footnote{In reality, mass is not a variable parameter, and the diatomics we consider are fictional. Nevertheless, the model parameter values were chosen to be consistent with realistic values found in diatomics, e.g. \cite{LI201378}} Meanwhile, $V(r)$ is given by the anharmonic Morse potential
\begin{equation}
    V(r)= D(1-e^{-\alpha (r-r_e)})^2-D.
    \label{MorsePotential}
\end{equation}

We consider a scenario where $r_e=1.3$ \AA \ is the equilibrium bond position, $D=37,000 \textrm{ cm}^{-1}$ is the potential well depth, and $\alpha = \frac{w_e}{2 r_e\sqrt{B_e D}}$ defines the width of the potential, where $w_e=3,000\textrm{ cm}^{-1}$ and $B_e = 11\textrm{ cm}^{-1}$. In addition, we assume the polarization of the field is aligned with the molecules' dipole moments, which are modeled in Padé-approximate form as in \cite{LI201378} by the function
\begin{equation}
    \mu(r) = M_0\frac{(1+x)^3}{1+\sum_{i=1}^4e_ix^i}\added{,}
\end{equation}
where $x \equiv \frac{r-r_e}{r_e}$ and the parameters are given by $M_0 = 0.5$ Debye, $e_1 = 2$, $e_2 = 2$, $e_3 = 2$, and $e_4 = 12$ (see Fig. \ref{Fig:Potentials} for plots of the Morse and dipole potentials). Using this model, the SSMC field is designed via Eq. \eqref{eq:molfield}, where the initial pump pulse is presumed to have the form
\begin{equation}
    E_p(t) = E_0 \sin^2(\pi t/T)\cos(w_e t)\added{,}
    \label{pumppulse}
\end{equation}
using $E_0 = 10^{-5}$ a.u.

Fig \ref{Fig:MolDiscFig} shows how SSMC performs in comparison with a naive approach, where a transform-limited pulse (as defined in Eq. (\ref{pumppulse})) is applied for the full time $t\in[0,n_sT]$. We choose this as a benchmark to compare the performance of SSMC against due to the fact that a transform-limited pulse is the shortest pulse with the highest peak intensity for a given bandwidth, and thus, these pulses will more efficiently generate non-linear effects (due to the peak intensity) compared to other pulses of equal total intensity. That is, all else being equal, a greater peak intensity should improve the distinguishability of different species' responses, as the nonlinearities will amplify the differences in the response of each species. Despite this, it is shown in Figs. \ref{Fig:MolDiscFig}(a)-(c) that as $T$, $n_s$, and $\Delta m$ are varied, the SSMC procedure leads to consistently lower values of $\textrm{cond}(\mathbf{A})$ than the naive approach by a few orders of magnitude. Importantly, we note that even when the intensity of the transform-limited pulses used in the naive implementation are increased by multiple orders of magnitude (keeping the SSMC fields unchanged), a clear separation in the performance of the two techniques persists.  

Clear trends are present in $\textrm{cond}(\mathbf{A})$ in Fig. \ref{Fig:MolDiscFig}(a), (b), and (c). For example, in Fig. \ref{Fig:MolDiscFig}(a), $\textrm{cond}(\mathbf{A})$ decreases with $T$ for both the SSMC and naive approach. This can be understood because the problem should become better conditioned with the addition of more data, which is what occurs when $T$ is increased. Meanwhile, the decrease in $\textrm{cond}(\mathbf{A})$ with $\Delta m$ in Fig. \ref{Fig:MolDiscFig}(c) is also to be expected, as increasing $\Delta m$ corresponds to making the different species in the mixture more distinct, thus making the mixture easier to characterize. Finally, the increase in $\textrm{cond}(\mathbf{A})$ with $n_s$ in Fig. \ref{Fig:MolDiscFig}(b) shows that as the complexity of a mixture increases, i.e. by adding more constituents, the task of characterizing the mixture becomes increasingly difficult. In fact, in Fig. \ref{Fig:MolDiscFig}(b) for SSMC it appears that the behavior of $\cond(\mathbf{A})$ is approximately linear for $n_s\geq 10$, while for the naive approach it is also approximately linear for $n_s\geq 13$. This linearity corresponds an exponential scaling of $\cond(\mathbf{A})$ with respect to $n_s$, and if these trends continue, one can expect the relative advantage of SSMC to remain constant as the number of mixture consituents increases.

In Fig. \ref{fig:EpsVsCond}, we report the relation between the condition of the problem, quantified by $\textrm{cond}(\mathbf{A})$ as per Eq. (\ref{Eq:CondA}), and the error in the solution, quantified by $\epsilon$ as per Eq. (\ref{Eq:Epsilon}). All of the data from Fig. \ref{Fig:MolDiscFig} is collected in Fig. \ref{fig:EpsVsCond}, such that each point marker in Fig. \ref{fig:EpsVsCond} has a corresponding point marker in Fig. \ref{Fig:MolDiscFig}. The filled point markers correspond to the errors obtained in an ideal, noise free manner, while the unfilled point markers are associated with a noisy implementation, where each element of the vector $\mathbf{R_{mix}}$ is infected by additive random noise, drawn from a Gaussian distribution with zero mean and a standard deviation of $0.001\norm{\mathbf{R_{mix}}}_\infty$. Note that the size of the noise variance is chosen rather arbitrarily, as the purpose of this addition is to assess the relative performance of SSMC and the naive approach, establish the condition number as a reliable proxy for error, and introduce a source of noise other than machine precision in the calculation of errors.

The rationale for adding noise to $\mathbf{R_{mix}}$ is as follows. First, within any real setting there will be errors from a number of sources. These may be measurement errors, as well as additional unknown species which will induce their own errors. The effect of these will be to add noises to the linear relationship in Eq. \eqref{lstsqs}:
\begin{equation}
   \left(\mathbf{A}+\mathbf{\Sigma}\right) \mathbf{\bar{y}} = \mathbf{R_{mix}}+\mathbf{\Gamma}, 
\end{equation}
where $\mathbf{\Sigma}$ represents an error in the response library matrix, and $\mathbf{\Gamma}$ an error in the observed mixture response. Note that we do not specify the distribution for these stochastic terms. Rearranging and expressing this equation with explicit indices, we have  
\begin{equation}
   \sum_{j} \mathbf{A}_{i,j} \mathbf{\bar{y}}_j = \mathbf{R_{mix}}_{,i}+\mathbf{\Gamma}_i +\sum_j\mathbf{\Sigma}_{i,j}\mathbf{\bar{y}}_j.
\end{equation}
The terms $\mathbf{\Gamma}_i +\sum_j\mathbf{\Sigma}_{i,j}\mathbf{\bar{y}}_j$ represent a sum of random variables, which by the central limit theorem will be well approximated as a single normally distributed random variable. Hence, the set of errors present both in the individual species' library response and the measured total response can be collected into a single Gaussian error on  $ \mathbf{R_{mix}}$. By representing this error as an additive noise to the mixture response, we are able to capture its effects while being agnostic as to its source. Furthermore, by assigning these errors to $\mathbf{R_{mix}}$, we are able to retain $\cond(\mathbf{A})$ as a deterministic figure of merit.

For both noiseless and noisy implementations depicted in Fig. \ref{fig:EpsVsCond}, the true solution $\mathbf{y}$ is taken to contain relative species concentrations drawn at random from a uniform distribution (and subsequently normalized). We note that there is a clear, approximately linear correlation between the condition number $\textrm{cond}(\mathbf{A})$ and the resulting error in the solution $\epsilon$, both when noise is added and when it is not. This suggests that the trends displayed in Fig. \ref{Fig:MolDiscFig}, showing values of $\textrm{cond}(\mathbf{A})$ that are consistently a few orders of magnitude lower for the SSMC procedure compared with the naive procedure, can be expected to correlate in practice to errors that are also significantly lower.

\section{Characterization of solid-state systems } \label{Sec:SolidState}

In order to demonstrate the flexibility of SSMC, we now apply it to a rather different set of model systems. Instead of molecular mixtures, we now consider a solid-state system, composed of non-interacting layers of different materials. In the same way as one determines relative concentrations of molecular species in a mixture, the relative density of each species in this layered material can also be determined using SSMC.

For the sake of simplicity, each layer will be described as a  1D Fermi-Hubbard model \cite{Tasaki1998}. Its dynamics display a number of interesting features, but the most relevant to the current discussion is that, like the molecular mixtures previously considered, this system also exhibits a highly non-linear response to driving \citep{Ghimire2012, Ghimire2011a, Murakami2018,Silva2018}. Furthermore, the tracking equations for the optical response of the Hubbard model are of an entirely different form to those derived in Sec. \ref{sec:molcharacterisation}. The choice of the Hubbard model is somewhat arbitrary, as these tracking equations for current apply to any system with an $n$-body potential dependent only on position. This provides a good test for the SSMC, as a way of checking that its effectiveness is not tied to a particular type of dynamical system. A full discussion of the tracking strategy required for many-body systems of this type (and its connection to the tracking equations used in Sec. \ref{sec:molcharacterisation}) may be found in Refs. \citep{mccaul2020controlling,mccaul2020driven}, but we present a brief recapitulation here. 

\subsection{Tracking Model}
We model our collection of systems using the Fermi-Hubbard model
\begin{align}
{H}^{(s)}\text{\ensuremath{\left(t\right)}}= & -t_{0} \sum_{j=1\atop\sigma=\uparrow, \downarrow}^{L} \text{\ensuremath{\left({\rm e}^{-i\Phi\left(t\right)}{c}_{j\sigma}^{\dagger}{c}_{j+1\sigma}+{\rm e}^{i\Phi\left(t\right)}{c}_{j+1\sigma}^{\dagger}{c}_{j\sigma}\right)}} +U^{(s)}\sum_{j=1}^{L}{c}_{j\uparrow}^{\dagger}{c}_{j\uparrow}{c}_{j\downarrow}^{\dagger}{c}_{j\downarrow}\added{,}
\label{eq:Hamiltonian}\end{align}
where the onsite interaction potential $U^{(s)}$ is used to parametrize different systems. The spin $\sigma$ fermionic annhilation operator for the $j$th site is given by $c_{j \sigma}$ and $t_0$ is the hopping parameter. Finally, the driving field  $E(t)$ enters the model through a P\added{e}ierls substitution \cite{feynman2011the}, $\Phi(t)=a\int^t_0 {\rm d}t^\prime E(t^\prime)$, where $a$ is the lattice constant. 

For these systems the optical response corresponds to the electronic current defined via a continuity equation for ${\rho_j}=\sum_{ \sigma=\uparrow, \downarrow}{c}_{j\sigma}^{\dagger} {c}_{j \sigma}$:
\begin{align}
    \frac{{\rm d}{\rho_j}}{{\rm d}t}&=\frac{1}{a}({R}_j-{R}_{j-1}),  \\
    {R}_j&=-iat_0\sum_{\sigma=\uparrow, \downarrow}\left({\rm e}^{-i\Phi\left(t\right)}{c}_{j\sigma}^{\dagger}{c}_{j+1\sigma}-{\rm h.c.}\right).
\end{align}
Summing over sites, we obtain the full response $R=\sum_j R_j$,
\begin{equation}
{R}=-iat_{0}\sum_{j=1\atop\sigma=\uparrow, \downarrow}^{L}\left({\rm e}^{-i\Phi\left(t\right)}{c}_{j\sigma}^{\dagger}{c}_{j+1\sigma}-{\rm h.c.}\right),\label{eq:currentoperator}
\end{equation}
with the response expectation given by 
\begin{equation}
    R^{(s)}(t)= \left\langle \psi^{(s)} (t) \left|{R}\right| \psi^{(s)}  (t) \right\rangle.
\end{equation}
Note that this expression for the response (and therefore the tracking equation derived from it) is identical for any system with a purely spatial potential and a dipolar coupling to the external field. If one defines the nearest neighbour expectation in a polar form
\begin{equation}
\left\langle \psi^{(s)}  (t) \left|\sum_{j=1\atop\sigma=\uparrow, \downarrow}^{L} {c}_{j\sigma}^{\dagger}{c}_{j+1\sigma}\right| \psi^{(s)}  (t) \right\rangle =K \left(\psi^{(s)} \right){\rm e}^{i\theta \left(\psi^{(s)} \right)}, \label{neighbourexpectation}
\end{equation}
then an explicit relationship between the response trajectory and the control field can be established:
\begin{equation}
    R^{(s)}(t)=-2at_0K(\psi^{(s)})\sin(\Phi(t)-\theta(\psi^{(s)})).
\end{equation}
Inverting this equation, one can find an expression for a ``tracking field'' $\Phi_T(t,\psi^{(s)})$, which produces the desired tracking response trajectory $R^{(s)}_T$ in system $\psi^{(s)}$:
\begin{align}
\Phi_T\left(t, \psi^{(s)} \right)&=\arcsin\left[-X(t,\psi^{(s)} )\right]+\theta\left(\psi^{(s)} \right),
\label{eq:phi_track} \\
   X(t,\psi^{(s)} ) & = \frac{R^{(s)} _T\left(t\right)}{2at_{0}K\left(\psi^{(s)} \right)}. 
\end{align}
Substituting this tracking field into Eq.~\eqref{eq:Hamiltonian}, we re-express the system dynamics with a ``tracking Hamiltonian'' ${H}^{(s)}_T (R^{(s)}_T(t),\psi^{(s)} )$ parametrised by the tracked response trajectory $R^{(s)}_T(t)$:
\begin{align}
{H}^{(s)} _{T}\left(R^{(s)}_T(t), \psi^{(s)} \right) & = \sum_{j=1\atop\sigma=\uparrow, \downarrow}^{L} \left[ P^{(s)} {\rm e}^{-i\theta \left(\psi^{(s)}  \right)}{c}_{j\sigma}^{\dagger}{c}_{j+1\sigma} + \mbox{h.c.} \right] \nonumber  + U^{(s)} \sum_{j=1}^{L}{c}_{j\uparrow}^{\dagger}{c}_{j\uparrow}{c}_{j\downarrow}^{\dagger}{c}_{j\downarrow}, \label{eq:trackingHamiltonian} \\
P^{(s)}  = -t_{0}&\left(\sqrt{1 - X^2(t,\psi^{(s)} )}+ i X(t,\psi^{(s)} )\right). 
\end{align}
Evolving with this tracking Hamiltonian is equivalent to generating dynamics with Eq.~\eqref{eq:Hamiltonian} when the driving field is chosen such that $\langle {R}\rangle =R^{(s)}_T (t)$. For SSMC, during the tracking phase for each species, we set $R^{(s)}_T (t)=0$. 

Finally, it is possible to show that the solutions for the tracking equations are unique (and therefore singularity free) provided certain constraints are satisfied. The interested reader may refer to Ref. \cite{mccaul2020controlling} for further details.

\subsection{Numerical illustrations}
Equipped with the tracking equations for the Hubbard model, we are able to implement SSMC and compare its performance to a transform-limited pulse. In this case we take a pump pulse given by
\begin{equation}
    \Phi_p(t)=a\frac{E_0}{\omega_0}\sin^2\left( \frac{\pi t}{T}\right)\sin(\omega_0t)\added{,}
    \label{eq:refphase}    
\end{equation}
where $E_0=10$MeV/cm is the electric field amplitude and $T$ is set as two periods of angular frequency $\omega_0=32.9$THz. The naive discriminating pulse will be a transform-limited pulse with the same functional form as the pump pulse, but running for the full time  $t\in[0,n_sT]$. For both the naive approach and SSMC, simulations are run for $L=10$ sites at half-filling (where the number of spin up and down electrons are equal, and the overall electron density is one per site), with each species initially in its ground state. We note that in general an arbitrary state (such as a Gibbs state) may be used, but given the energy gap in the model considered is much greater than thermal fluctuations at room temperature, the ground state is an convenient choice for the initial preparation.

\begin{figure}[t]
    \centering
        \includegraphics[width=0.6\columnwidth]{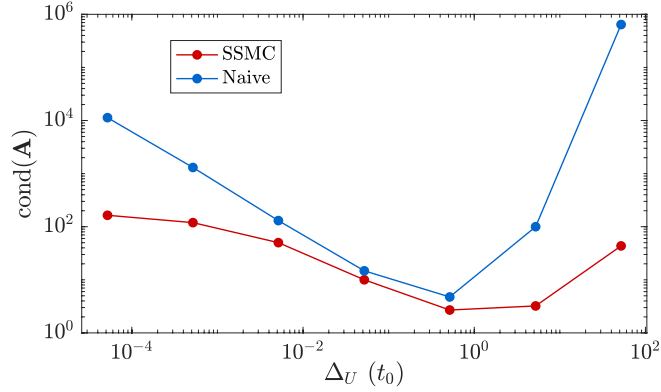}
    \caption{Performance of optical discrimination vs naive approach when $\Delta_U$ is varied in a two species mixture. Here $U^{(0)}=t_0$ and $U^{(1)} = U^{(0)} +\Delta_U$. \label{fig:2speciesresults}}
\end{figure}

We first examine two systems in a mixture where $t_0$ is constant but the two species have onsite repulsions $U^{(0)}$ and $U^{(1)}$. We consider the relative accuracy of both the naive method and optical discrimination as we vary the distance in the two species' onsite repulsions, i.e. $U^{(1)} = U^{(0)} +\Delta_U$. Fig. \ref{fig:2speciesresults} shows that over a wide range of $\Delta_U$, SSMC outperforms the naive approach. This increase in efficiency is of particular value when $\Delta_U$ is small, and the systems are not easily distinguishable from each other. While the precise dynamics of each system will depend on both the choice of $U^{(0)}$ and $\Delta_U$, SSMC retains its advantage over the naive approach regardles of the choice of $U^{(0)}$.

\begin{figure}[t]
    \centering
        \includegraphics[width=0.4\columnwidth]{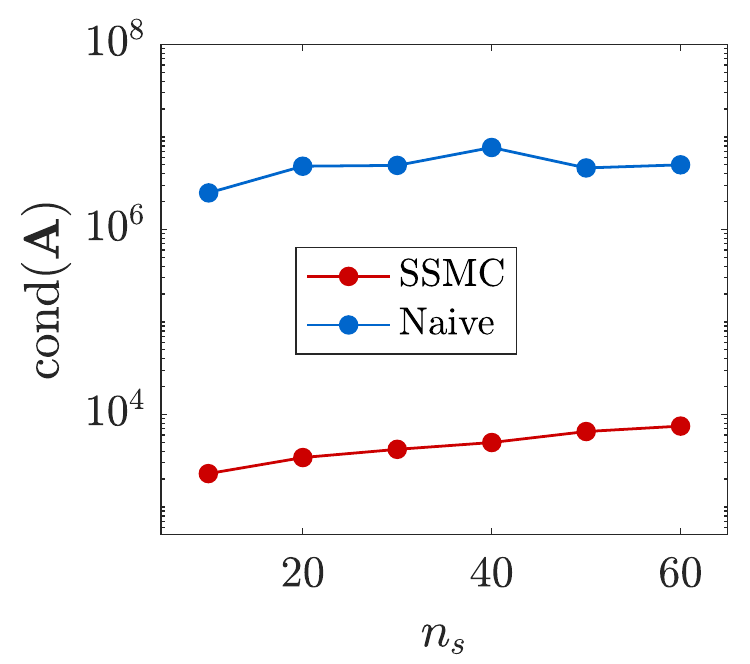}
    \caption{Performance of optical discrimination vs naive approach when the number of species $n_s$ is varied. Here, $U^{(0)}=t_0$ and $U^{(n_s)}=(1+10^{-2})t_0$, with each of the $n_s$ species being equally spaced in $U$.
    \label{fig:Nspeciesresults}}
\end{figure}

The accuracy gain from using SSMC can be seen more clearly when generalising to the $n_s$ species case, shown in  Fig. \ref{fig:Nspeciesresults}. Here we find qualitatively identitical results to those shown in Fig.~\ref{Fig:MolDiscFig}, where there is a clear advantage to using SSMC, and that the relative performance between the two methods remains roughly constant as one increases $n_s$. Furthermore, this consistency with the results of Sec. \ref{sec:molcharacterisation} extends to the relationship between the condition number and error $\epsilon$.

\section{Conclusion} \label{Sec:Conclusion}

In this article, we have introduced SSMC as a new approach for characterizing mixtures of non-interacting components, which uses a tailored field designed to sequentially suppress the optical responses of the mixture components. The duration of this SSMC field scales linearly in the number of components.  We have derived expressions for the SSMC fields needed to carry out this procedure using quantum tracking control principles, and have numerically studied applications of SSMC to mixtures of diatomic molecules in the gas phase, as well as to solid-state systems. In each case, we compared the performance of SSMC with a naive approach involving the application of a simple, transform-limited pulse to the mixture. As illustrated in Figs. \ref{Fig:MolDiscFig}, \ref{fig:2speciesresults}, and \ref{fig:Nspeciesresults}, we found that the SSMC procedure leads to characterization problems that are significantly better conditioned (i.e., as quantified by condition numbers that are consistently orders of magnitude lower, compared with those obtained utilizing a transform limited pulse). We further found that the condition number can reliably serve as a proxy for the error in the solution, as demonstrated by the clear correlation between the two plotted in Fig. \ref{fig:EpsVsCond}. 

The numerical analyses presented in Sections \ref{sec:molcharacterisation} and \ref{Sec:SolidState} studied the performance of SSMC with respect to different variables, such as the number of species in the mixture $n_s$ and $\Delta m$ and $\Delta_U$, which represent the similarity of species in mixtures of molecular and solid state constituents, respectively. For each study, SSMC fields $E(t)$ were found by concatenating a transform-limited pump pulse with a series of different pulses designed to sequentially suppress the optical responses of the species in the mixture. For each case presented here, the suppression order was chosen such that species were suppressed in ascending order according to their value of $m^{(s)}$ or $U^{(s)}$. We remark that this choice was arbitrary, and further simulations with different orderings were performed to assess the impact of the suppression order on the performance. For both example systems, it was found that the suppression order did not significantly affect the values obtained for $\cond(\mathbf{A})$ or $\epsilon$. 

We also remark that although the examples in Section \ref{sec:molcharacterisation} considered molecular systems modeled in the Born-Oppenheimer approximation, the SSMC tracking control procedure is far more general, and can be extended in a straightforward manner to systems outside of the Born-Oppenheimer approximation. In addition, both of the examples considered in Sections \ref{sec:molcharacterisation} and \ref{Sec:SolidState} involve the application of the SSMC procedure to the characterization of mixtures of closed quantum systems undergoing unitary dynamics. Again, the SSMC procedure is far more general, and can be extended to address open quantum systems with dissipative dynamics as well \cite{PhysRevLett.118.083201}. A full derivation treating these more general cases is provided in Appendix \ref{Appendix}.

The SSMC procedure we have introduced stands as an alternative to the naive approach depicted in Fig. \ref{Fig:NaiveProcedure}, and also to other mixture characterization methods, such as ODD. Regarding the latter, we remark that the differences between ODD and SSMC make a comprehensive comparison of the performance of the two approaches difficult. Namely, ODD is an experimental procedure for iteratively identifying pulses for mixture characterization, while SSMC involves numerical pulse design. Furthermore, there is flexibility in selecting the objective function used in ODD to optimize the pulse in the laboratory, and the performance of ODD will depend on this choice. For example, if the objective function is selected to match the goals of SSMC (i.e., to sequentially turn off the optical responses of different mixture components), then we expect that ODD could achieve similar performance to SSMC. However, when alternate objective functions are chosen for ODD (e.g., to reflect measurable quantities in experiments), then it is difficult to speculate about how ODD would perform compared with SSMC.

In the laboratory, the capabilities of laser sources (e.g. bandwidth, intensity) continue to rapidly improve \cite{elu2020seven,lesko2020six}, and new techniques for accelerating data acquisition \cite{PhysRevApplied.15.024032} and characterizing the high frequency non-linear part of the response field \cite{Sommer2016} (necessary for any implementation of tracking control) have recently been demonstrated. These developments suggest that successful experimental implementations of SSMC for a variety of mixture types may be feasible today, or in the near future. Furthermore, the SSMC procedure can also be modified to better suit the capabilities of existing laser technologies. For example, it has been shown that the tracking control methods on which SSMC is based produce control fields which can be well approximated by only a \emph{few distinct frequencies} \cite{mccaul2020controlling}; additionally, the method used to construct them is itself robust to noise \cite{PhysRevLett.118.083201}. One would therefore expect to be able to construct an approximate library pulse for discrimination, and provided the severity of the approximation is small compared to other sources of error, this pulse would still provide significant improvements in accuracy over an unshaped or transform-limited pulse. 

It is also worth remarking that in practice, the computational challenge of determining the SSMC fields for mixtures whose constituents are complex quantum systems can be significant, given that the complexity of the simulations is exponential in the number of degrees of freedom of the constituents. To address this, numerous quantum dynamics approximation methods such as time-dependent density functional theory \cite{PhysRevLett.109.153603}, (multiconfigurational) time-dependent Hartree(-Fock) \cite{Messina1996,Schroder2008,Mundt_2009}, and tensor network methods such as time-dependent density matrix renormalization group \cite{PhysRevLett.106.190501} have been developed and shown to yield impressive performance for different classes of systems at reduced computational cost, and could therefore serve as useful tools for use in the numerical SSMC field design procedure. Given that any error in the relevant tracking portion of the evolution will resemble a fluctuation around zero response (and thus resembles an error of the type considered in Sec. \ref{sec:molcharacterisation}), we expect that errors stemming from the use of these approximation frameworks will be tolerable, and still lead to SSMC fields that can be used to successfully characterize mixtures. Furthermore, in the future it may even be possible to leverage quantum computing devices in order to facilitate the design of fields for controlling quantum systems \cite{magann2020digital}.

Finally, we note that numerical tracking control is not the only method through which one can attain the essential goal - i.e. the sequential extinguishing of each species' optical response. Deep learning networks have previously been employed to experimentally determine the input field required to generate a desired response, in a manner that is robust to noise. This technique could easily be applied to the SSMC scheme for the calculation of the library discrimination pulse \cite{lohani_dispersion_2019}.

\begin{acknowledgments}

A.B.M. acknowledges support from the U.S. Department of Energy, Office of Science, Office of Advanced Scientific Computing Research, Department of Energy Computational Science Graduate Fellowship (grant DE-FG02-97ER25308), as well as helpful discussions with C. Arenz. H.A.R. acknowledges support from the U.S. Army Research Office grant numbers (W911NF-16-1-0014 and W911NF-19-1-0382) for theoretical foundations, and the U.S. Department of Energy grant number (DE-FG02-02ER15344 ) for formulation of applications. G.M. and D.I.B. are supported by Air Force Office of Scientific Research (AFOSR) Young Investigator Research Program (grant FA9550-16-1-0254) and the Army Research Office (ARO) (grant W911NF-19-1-0377). The views and conclusions contained in this document are those of the authors and should not be interpreted as representing the official policies, either expressed or implied, of AFOSR, ARO, or the U.S. Government. The U.S. Government is authorized to reproduce and distribute reprints for Government purposes notwithstanding any copyright notation herein.

This report was prepared as an account of work sponsored by an agency of the United States Government. Neither the United States Government nor any agency thereof, nor any of their employees, makes any warranty, express or implied, or assumes any legal liability or responsibility for the accuracy, completeness, or usefulness of any information, apparatus, product, or process disclosed, or represents that its use would not infringe privately owned rights. Reference herein to any specific commercial product, process, or service by trade name, trademark, manufacturer, or otherwise does not necessarily constitute or imply its endorsement, recommendation, or favoring by the United States Government or any agency thereof. The views and opinions of authors expressed herein do not necessarily state or reflect those of the United States Government or any agency thereof. 

Sandia National Laboratories is a multimission laboratory managed and operated by National Technology \& Engineering Solutions of Sandia, LLC, a wholly owned subsidiary of Honeywell International Inc., for the U.S. Department of Energy's National Nuclear Security Administration under contract DE-NA0003525. This paper describes objective technical results and analysis. Any subjective views or opinions that might be expressed in the paper do not necessarily represent the views of the U.S. Department of Energy or the United States Government. SAND2022-0139 J.

\end{acknowledgments}

\section*{Author Contributions}
{A.B.M and G.M. contributed equally to this work}. 

\bibliographystyle{plainnat}
\bibliography{bib}

\providecommand{\noopsort}[1]{}\providecommand{\singleletter}[1]{#1}%
\begin{thebibliography}{47}
\providecommand{\natexlab}[1]{#1}
\providecommand{\url}[1]{\texttt{#1}}
\expandafter\ifx\csname urlstyle\endcsname\relax
  \providecommand{\doi}[1]{doi: #1}\else
  \providecommand{\doi}{doi: \begingroup \urlstyle{rm}\Url}\fi

\bibitem[Adhikari et~al.(2021)Adhikari, Cortes, Wen, Panuganti, Gosztola,
  Schaller, Wiederrecht, and Gray]{PhysRevApplied.15.024032}
Sushovit Adhikari, Cristian~L. Cortes, Xiewen Wen, Shobhana Panuganti, David~J.
  Gosztola, Richard~D. Schaller, Gary~P. Wiederrecht, and Stephen~K. Gray.
\newblock Accelerating ultrafast spectroscopy with compressive sensing.
\newblock \emph{Phys. Rev. Applied}, 15:\penalty0 024032, 2021.
\newblock \doi{10.1103/PhysRevApplied.15.024032}.

\bibitem[Bagwell and Adams(1993)]{bagwell1993fluorescence}
C~Bruce Bagwell and Earl~G Adams.
\newblock Fluorescence spectral overlap compensation for any number of flow
  cytometry parameters.
\newblock \emph{Ann. N. Y. Acad. Sci.}, 677\penalty0 (1):\penalty0 167--184,
  1993.
\newblock \doi{10.1111/j.1749-6632.1993.tb38775.x}.

\bibitem[Brixner et~al.(2001)Brixner, Damrauer, Niklaus, and
  Gerber]{brixner2001photoselective}
T~Brixner, NH~Damrauer, P~Niklaus, and G~Gerber.
\newblock Photoselective adaptive femtosecond quantum control in the liquid
  phase.
\newblock \emph{Nature}, 414\penalty0 (6859):\penalty0 57--60, 2001.
\newblock \doi{10.1038/35102037}.

\bibitem[Campos et~al.(2017)Campos, Bondar, Cabrera, and
  Rabitz]{PhysRevLett.118.083201}
Andre~G. Campos, Denys~I. Bondar, Renan Cabrera, and Herschel~A. Rabitz.
\newblock How to make distinct dynamical systems appear spectrally identical.
\newblock \emph{Phys. Rev. Lett.}, 118:\penalty0 083201, 2017.
\newblock \doi{10.1103/PhysRevLett.118.083201}.

\bibitem[Castro et~al.(2012)Castro, Werschnik, and
  Gross]{PhysRevLett.109.153603}
A.~Castro, J.~Werschnik, and E.~K.~U. Gross.
\newblock Controlling the dynamics of many-electron systems from first
  principles: A combination of optimal control and time-dependent
  density-functional theory.
\newblock \emph{Phys. Rev. Lett.}, 109:\penalty0 153603, 2012.
\newblock \doi{10.1103/PhysRevLett.109.153603}.

\bibitem[Chen et~al.(1995)Chen, Gross, Ramakrishna, Rabitz, and
  Mease]{Chen1995}
Yu~Chen, Peter Gross, Viswanath Ramakrishna, Herschel Rabitz, and Kenneth
  Mease.
\newblock {Competitive tracking of molecular objectives described by quantum
  mechanics}.
\newblock \emph{J. Chem. Phys.}, 102\penalty0 (8001), 1995.
\newblock \doi{10.1063/1.468998}.

\bibitem[Chen et~al.(1997)Chen, Gross, Ramakrishna, Rabitz, Mease, and
  Singh]{Chen1997}
Yu~Chen, Peter Gross, Viswanath Ramakrishna, Herschel Rabitz, Kenneth Mease,
  and Harjinder Singh.
\newblock {Control of Classical Regime Molecular Objectives -Applications of
  Tracking and Variations on the Theme*}.
\newblock \emph{Automatica}, 33\penalty0 (9):\penalty0 1617--1633, 1997.
\newblock \doi{10.1016/S0005-1098(97)00077-0}.

\bibitem[Doria et~al.(2011)Doria, Calarco, and
  Montangero]{PhysRevLett.106.190501}
Patrick Doria, Tommaso Calarco, and Simone Montangero.
\newblock Optimal control technique for many-body quantum dynamics.
\newblock \emph{Phys. Rev. Lett.}, 106:\penalty0 190501, 2011.
\newblock \doi{10.1103/PhysRevLett.106.190501}.

\bibitem[Elu et~al.(2020)Elu, Maidment, Vamos, Tani, Novoa, Frosz, Badikov,
  Badikov, Petrov, Russell, et~al.]{elu2020seven}
Ugaitz Elu, Luke Maidment, Lenard Vamos, Francesco Tani, David Novoa, Michael~H
  Frosz, Valeriy Badikov, Dmitrii Badikov, Valentin Petrov, Philip St~J
  Russell, et~al.
\newblock Seven-octave high-brightness and carrier-envelope-phase-stable light
  source.
\newblock \emph{Nat. Photonics}, pages 1--4, 2020.
\newblock \doi{10.1038/s41566-020-00735-1}.

\bibitem[Feng et~al.(2000)Feng, Mellor, Bernstein, Keller-Peck, Nguyen,
  Wallace, Nerbonne, Lichtman, and Sanes]{feng2000imaging}
Guoping Feng, Rebecca~H Mellor, Michael Bernstein, Cynthia Keller-Peck, Quyen~T
  Nguyen, Mia Wallace, Jeanne~M Nerbonne, Jeff~W Lichtman, and Joshua~R Sanes.
\newblock Imaging neuronal subsets in transgenic mice expressing multiple
  spectral variants of gfp.
\newblock \emph{Neuron}, 28\penalty0 (1):\penalty0 41--51, 2000.
\newblock \doi{10.1016/S0896-6273(00)00084-2}.

\bibitem[Feynman(2011)]{feynman2011the}
Richard Feynman.
\newblock \emph{The Feynman lectures on physics}.
\newblock Basic Books, a member of the Perseus Books Group, New York, 2011.
\newblock ISBN 0465023827.

\bibitem[Ghimire et~al.(2011)Ghimire, Dichiara, Sistrunk, Agostini, Dimauro,
  and Reis]{Ghimire2011a}
Shambhu Ghimire, Anthony~D. Dichiara, Emily Sistrunk, Pierre Agostini, Louis~F.
  Dimauro, and David~A. Reis.
\newblock {Observation of high-order harmonic generation in a bulk crystal}.
\newblock \emph{Nat. Phys.}, 7\penalty0 (2):\penalty0 138--141, 2011.
\newblock \doi{10.1038/nphys1847}.

\bibitem[Ghimire et~al.(2012)Ghimire, Dichiara, Sistrunk, Ndabashimiye,
  Szafruga, Mohammad, Agostini, Dimauro, and Reis]{Ghimire2012}
Shambhu Ghimire, Anthony~D. Dichiara, Emily Sistrunk, Georges Ndabashimiye,
  Urszula~B. Szafruga, Anis Mohammad, Pierre Agostini, Louis~F. Dimauro, and
  David~A. Reis.
\newblock {Generation and propagation of high-order harmonics in crystals}.
\newblock \emph{Phys. Rev. A}, 85\penalty0 (4):\penalty0 043836, 2012.
\newblock \doi{10.1103/PhysRevA.85.043836}.

\bibitem[{Goun} et~al.(2016){Goun}, {Bondar}, {Er}, {Quine}, and
  {Rabitz}]{2016NatSR...625827G}
Alexei {Goun}, Denys~I. {Bondar}, Ali~O. {Er}, Zachary {Quine}, and Herschel~A.
  {Rabitz}.
\newblock {Photonic reagents for concentration measurement of fluorescent
  proteins with overlapping spectra}.
\newblock \emph{Sci. Rep.}, 6:\penalty0 25827, 2016.
\newblock \doi{10.1038/srep25827}.

\bibitem[Gross et~al.(1993)Gross, Singh, Rabitz, Mease, and
  Huang]{gross1993inverse}
Peter Gross, Harjinder Singh, Herschel Rabitz, Kenneth Mease, and GM~Huang.
\newblock Inverse quantum-mechanical control: A means for design and a test of
  intuition.
\newblock \emph{Phys. Rev. A}, 47\penalty0 (6):\penalty0 4593, 1993.
\newblock \doi{10.1103/PhysRevA.47.4593}.

\bibitem[Lesko et~al.(2021)Lesko, Timmers, Xing, Kowligy, Lind, and
  Diddams]{lesko2020six}
Daniel~MB Lesko, Henry Timmers, Sida Xing, Abijith Kowligy, Alexander~J Lind,
  and Scott~A Diddams.
\newblock A six-octave optical frequency comb from a scalable few-cycle erbium
  fibre laser.
\newblock \emph{Nat. Photonics}, 15\penalty0 (4):\penalty0 281--286, 2021.
\newblock \doi{10.1038/s41566-021-00778-y}.

\bibitem[Li et~al.(2002)Li, Turinici, Ramakrishna, and Rabitz]{oddrabitz}
Baiqing Li, Gabriel Turinici, Viswanath Ramakrishna, and Herschel Rabitz.
\newblock Optimal dynamic discrimination of similar molecules through quantum
  learning control.
\newblock \emph{J. Phys. Chem. B}, 106\penalty0 (33):\penalty0 8125--8131,
  2002.
\newblock \doi{10.1021/jp0204657}.

\bibitem[Li et~al.(2005)Li, Rabitz, and Wolf]{li2005optimal}
Baiqing Li, Herschel Rabitz, and Jean-Pierre Wolf.
\newblock Optimal dynamic discrimination of similar quantum systems with time
  series data.
\newblock \emph{J. Chem. Phys}, 122\penalty0 (15):\penalty0 154103, 2005.
\newblock \doi{10.1063/1.1883170}.

\bibitem[Li et~al.(2013)Li, Gordon, Roy, Hajigeorgiou, Coxon, Bernath, and
  Rothman]{LI201378}
Gang Li, Iouli~E. Gordon, Robert J.~Le Roy, Photos~G. Hajigeorgiou, John~A.
  Coxon, Peter~F. Bernath, and Laurence~S. Rothman.
\newblock Reference spectroscopic data for hydrogen halides. part i:
  Construction and validation of the ro-vibrational dipole moment functions.
\newblock \emph{J. Quant. Spectrosc. Radiat. Transf.}, 121:\penalty0 78 -- 90,
  2013.
\newblock \doi{10.1016/j.jqsrt.2013.02.005}.

\bibitem[Lichtman and Conchello(2005)]{lichtman2005fluorescence}
Jeff~W Lichtman and Jos{\'e}-Angel Conchello.
\newblock Fluorescence microscopy.
\newblock \emph{Nat. Methods}, 2\penalty0 (12):\penalty0 910--919, 2005.
\newblock \doi{10.1038/nmeth817}.

\bibitem[Lidar and Schneider(2005)]{Lidar2004}
Daniel~A Lidar and Sara Schneider.
\newblock Stabilizing qubit coherence via tracking-control.
\newblock \emph{Quantum Inf. Comput.}, 5\penalty0 (4), 2005.
\newblock URL \url{https://dl.acm.org/doi/10.5555/2011645.2011651}.

\bibitem[Lohani et~al.(2019)Lohani, Knutson, Zhang, and
  Glasser]{lohani_dispersion_2019}
Sanjaya Lohani, Erin~M. Knutson, Wenlei Zhang, and Ryan~T. Glasser.
\newblock Dispersion characterization and pulse prediction with machine
  learning.
\newblock \emph{OSA Continuum}, 2\penalty0 (12):\penalty0 3438, 2019.
\newblock \doi{10.1364/OSAC.2.003438}.

\bibitem[Magann et~al.(2018)Magann, Ho, and Rabitz]{PhysRevA.98.043429}
Alicia Magann, Tak-San Ho, and Herschel Rabitz.
\newblock Singularity-free quantum tracking control of molecular rotor
  orientation.
\newblock \emph{Phys. Rev. A}, 98:\penalty0 043429, 2018.
\newblock \doi{10.1103/PhysRevA.98.043429}.

\bibitem[Magann et~al.(2021)Magann, Grace, Rabitz, and
  Sarovar]{magann2020digital}
Alicia~B. Magann, Matthew~D. Grace, Herschel~A. Rabitz, and Mohan Sarovar.
\newblock Digital quantum simulation of molecular dynamics and control.
\newblock \emph{Phys. Rev. Research}, 3:\penalty0 023165, Jun 2021.
\newblock \doi{10.1103/PhysRevResearch.3.023165}.

\bibitem[McCaul et~al.(2020{\natexlab{a}})McCaul, Orthodoxou, Jacobs, Booth,
  and Bondar]{mccaul2020controlling}
Gerard McCaul, Christopher Orthodoxou, Kurt Jacobs, George~H Booth, and Denys~I
  Bondar.
\newblock Controlling arbitrary observables in correlated many-body systems.
\newblock \emph{Phys. Rev. A}, 101\penalty0 (5):\penalty0 053408,
  2020{\natexlab{a}}.
\newblock \doi{10.1103/PhysRevA.101.053408}.

\bibitem[McCaul et~al.(2020{\natexlab{b}})McCaul, Orthodoxou, Jacobs, Booth,
  and Bondar]{mccaul2020driven}
Gerard McCaul, Christopher Orthodoxou, Kurt Jacobs, George~H Booth, and Denys~I
  Bondar.
\newblock Driven imposters: Controlling expectations in many-body systems.
\newblock \emph{Phys. Rev. Lett.}, 124\penalty0 (18):\penalty0 183201,
  2020{\natexlab{b}}.
\newblock \doi{10.1103/PhysRevLett.124.183201}.

\bibitem[Medley et~al.(2005)Medley, Drake, Tomasini, Rogers, and
  Tan]{medley2005simultaneous}
Colin~D Medley, Timothy~J Drake, Jeffrey~M Tomasini, Richard~J Rogers, and
  Weihong Tan.
\newblock Simultaneous monitoring of the expression of multiple genes inside of
  single breast carcinoma cells.
\newblock \emph{Anal. Chem.}, 77\penalty0 (15):\penalty0 4713--4718, 2005.
\newblock \doi{10.1021/ac050881y}.

\bibitem[Messina et~al.(1996)Messina, Wilson, and Krause]{Messina1996}
Michael Messina, Kent~R Wilson, and Jeffrey~L Krause.
\newblock Quantum control of multidimensional systems: Implementation within
  the time?dependent hartree approximation.
\newblock \emph{J. Chem. Phys.}, 104:\penalty0 173, 1996.
\newblock \doi{10.1063/1.470887}.

\bibitem[Mitra and Rabitz(2003)]{mitra2003identifying}
Abhra Mitra and Herschel Rabitz.
\newblock Identifying mechanisms in the control of quantum dynamics through
  hamiltonian encoding.
\newblock \emph{Phys. Rev. A}, 67\penalty0 (3):\penalty0 033407, 2003.
\newblock \doi{10.1103/PhysRevA.67.033407}.

\bibitem[Mitra and Rabitz(2004)]{mitra2004mechanistic}
Abhra Mitra and Herschel Rabitz.
\newblock Mechanistic analysis of optimal dynamic discrimination of similar
  quantum systems.
\newblock \emph{J. Phys. Chem. A}, 108\penalty0 (21):\penalty0 4778--4785,
  2004.
\newblock \doi{10.1021/jp0495390}.

\bibitem[Mitra and Rabitz(2006)]{mitra2006quantum}
Abhra Mitra and Herschel Rabitz.
\newblock Quantum control mechanism analysis through field based hamiltonian
  encoding.
\newblock \emph{J. Chem. Phys.}, 125\penalty0 (19):\penalty0 194107, 2006.
\newblock \doi{10.1063/1.2371079}.

\bibitem[Morse(1929)]{PhysRev.34.57}
Philip~M. Morse.
\newblock Diatomic molecules according to the wave mechanics. ii. vibrational
  levels.
\newblock \emph{Phys. Rev.}, 34:\penalty0 57--64, 1929.
\newblock \doi{10.1103/PhysRev.34.57}.

\bibitem[Mundt and Tannor(2009)]{Mundt_2009}
Michael Mundt and David~J Tannor.
\newblock Optimal control of interacting particles: a multi-configuration
  time-dependent hartree{\textendash}fock approach.
\newblock \emph{New J. Phys}, 11\penalty0 (10):\penalty0 105038, 2009.
\newblock \doi{10.1088/1367-2630/11/10/105038}.

\bibitem[Murakami et~al.(2018)Murakami, Eckstein, and Werner]{Murakami2018}
Yuta Murakami, Martin Eckstein, and Philipp Werner.
\newblock {High-Harmonic Generation in Mott Insulators}.
\newblock \emph{Phys. Rev. Lett.}, 121\penalty0 (5):\penalty0 057405, 2018.
\newblock \doi{10.1103/PhysRevLett.121.057405}.

\bibitem[Perfetto et~al.(2004)Perfetto, Chattopadhyay, and
  Roederer]{perfetto2004seventeen}
Stephen~P Perfetto, Pratip~K Chattopadhyay, and Mario Roederer.
\newblock Seventeen-colour flow cytometry: unravelling the immune system.
\newblock \emph{Nat. Rev. Immunol.}, 4\penalty0 (8):\penalty0 648--655, 2004.
\newblock \doi{10.1038/nri1416}.

\bibitem[Rondi et~al.(2012)Rondi, Bonacina, Trisorio, Hauri, and
  Wolf]{rondi2012coherent}
Ariana Rondi, Luigi Bonacina, A~Trisorio, C~Hauri, and J-P Wolf.
\newblock Coherent manipulation of free amino acids fluorescence.
\newblock \emph{Phys. Chem. Chem. Phys.}, 14\penalty0 (26):\penalty0
  9317--9322, 2012.
\newblock \doi{10.1039/C2CP23357F}.

\bibitem[Roslund et~al.(2011)Roslund, Roth, Guyon, Boutou, Courvoisier, Wolf,
  and Rabitz]{roslund2011resolution}
Jonathan Roslund, Matthias Roth, Laurent Guyon, V{\'e}ronique Boutou, Francois
  Courvoisier, Jean-Pierre Wolf, and Herschel Rabitz.
\newblock Resolution of strongly competitive product channels with optimal
  dynamic discrimination: Application to flavins.
\newblock \emph{J. Chem. Phys.}, 134\penalty0 (3):\penalty0 034511, 2011.
\newblock \doi{10.1063/1.3518751}.

\bibitem[Roth et~al.(2009)Roth, Guyon, Roslund, Boutou, Courvoisier, Wolf, and
  Rabitz]{PhysRevLett.102.253001}
Matthias Roth, Laurent Guyon, Jonathan Roslund, V\'eronique Boutou, Francois
  Courvoisier, Jean-Pierre Wolf, and Herschel Rabitz.
\newblock Quantum control of tightly competitive product channels.
\newblock \emph{Phys. Rev. Lett.}, 102:\penalty0 253001, 2009.
\newblock \doi{10.1103/PhysRevLett.102.253001}.

\bibitem[Schr{\"o}der et~al.(2008)Schr{\"o}der, Carre{\'o}n-Macedo, and
  Brown]{Schroder2008}
Markus Schr{\"o}der, Jos{\'e}-Luis Carre{\'o}n-Macedo, and Alex Brown.
\newblock Implementation of an iterative algorithm for optimal control of
  molecular dynamics into mctdh.
\newblock \emph{Phys. Chem. Chem. Phys.}, 10:\penalty0 850, 2008.
\newblock \doi{10.1039/B714821F}.

\bibitem[Silva et~al.(2018)Silva, Blinov, Rubtsov, Smirnova, and
  Ivanov]{Silva2018}
R.~E.F. Silva, Igor~V. Blinov, Alexey~N. Rubtsov, O.~Smirnova, and M.~Ivanov.
\newblock {High-harmonic spectroscopy of ultrafast many-body dynamics in
  strongly correlated systems}.
\newblock \emph{Nat. Photonics}, 12\penalty0 (5):\penalty0 266--270, 2018.
\newblock \doi{10.1038/s41566-018-0129-0}.

\bibitem[Sommer et~al.(2016)Sommer, Bothschafter, Sato, Jakubeit, Latka,
  Razskazovskaya, Fattahi, Jobst, Schweinberger, Shirvanyan, Yakovlev,
  Kienberger, Yabana, Karpowicz, Schultze, and Krausz]{Sommer2016}
A~Sommer, E~M Bothschafter, S~A Sato, C~Jakubeit, T~Latka, O~Razskazovskaya,
  H~Fattahi, M~Jobst, W~Schweinberger, V~Shirvanyan, V~S Yakovlev,
  R~Kienberger, K~Yabana, N~Karpowicz, M~Schultze, and F~Krausz.
\newblock {Attosecond nonlinear polarization and light–matter energy transfer
  in solids}.
\newblock 534\penalty0 (7605):\penalty0 86--90, 2016.
\newblock \doi{10.1038/nature17650}.

\bibitem[Speicher et~al.(1996)Speicher, Ballard, and
  Ward]{speicher1996karyotyping}
Michael~R Speicher, Stephen~Gwyn Ballard, and David~C Ward.
\newblock Karyotyping human chromosomes by combinatorial multi-fluor fish.
\newblock \emph{Nat. Genet.}, 12\penalty0 (4):\penalty0 368--375, 1996.
\newblock \doi{10.1038/ng0496-368}.

\bibitem[Stapelfeldt and Seideman(2003)]{RevModPhys.75.543}
Henrik Stapelfeldt and Tamar Seideman.
\newblock Colloquium: Aligning molecules with strong laser pulses.
\newblock \emph{Rev. Mod. Phys.}, 75:\penalty0 543--557, Apr 2003.
\newblock \doi{10.1103/RevModPhys.75.543}.

\bibitem[Tasaki(1998)]{Tasaki1998}
Hal Tasaki.
\newblock {The Hubbard model - An introduction and selected rigorous results}.
\newblock \emph{J. Phys. Cond. Mat.}, 10\penalty0 (20):\penalty0 4353--4378,
  1998.
\newblock \doi{10.1088/0953-8984/10/20/004}.

\bibitem[Wilson(2000)]{wilson2000encyclopedia}
Ian Wilson.
\newblock \emph{Encyclopedia of separation science}.
\newblock Academic Press, San Diego, 2000.
\newblock ISBN 0122267702.

\bibitem[Zhu and Rabitz(2003)]{doi:10.1063/1.1582847}
Wusheng Zhu and Herschel Rabitz.
\newblock Quantum control design via adaptive tracking.
\newblock \emph{J. Chem. Phys.}, 119\penalty0 (7):\penalty0 3619--3625, 2003.
\newblock \doi{10.1063/1.1582847}.

\bibitem[Zhu et~al.(1999)Zhu, Smit, and Rabitz]{doi:10.1063/1.477857}
Wusheng Zhu, Martina Smit, and Herschel Rabitz.
\newblock Managing singular behavior in the tracking control of quantum
  dynamical observables.
\newblock \emph{J. Chem. Phys.}, 110\penalty0 (4):\penalty0 1905--1915, 1999.
\newblock \doi{10.1063/1.477857}.

\end{thebibliography}

\appendix

\section{First principles derivation} \label{Appendix}

In this section, we consider a derivation for the SSMC equations for a quantum mixture based on first principles. 
We begin by considering a component of a quantum mixture which is composed of $N_n^{(s)}$ nuclei and $N_e^{(s)}$ electrons, where the subscripts distinguish between nuclei and electrons and the superscripts label the mixture constituent they belong to. For systems undergoing dissipative dynamics, we may model the dynamics of the state $\rho^{(s)}(t)$ of the mixture component by the dynamical equation
\begin{equation}
\begin{aligned}
\frac{d}{dt}\rho^{(s)}(t)&=\mathcal{L}^{(s)}(\rho^{(s)}(t))\\
&=-i[H^{(s)}(t),\rho^{(s)}(t)]+\mathcal{D}^{(s)}(\rho^{(s)}(t)),
\label{DynamicalEq}
\end{aligned}
\end{equation}
which is of Lindblad form, where the dissipator $\mathcal{D}^{(s)}(\cdot)$ describes the interactions between $\rho^{(s)}(t)$ and its environment, given by
\begin{equation}
\begin{aligned}
    \mathcal{D}^{(s)}(\rho^{(s)}(t))&=\sum_{j,k}\gamma_{jk}^{(s)}\Big(A_{jk}^{(s)}\rho^{(s)}(t)A_{jk}^{(s)\dagger}-\frac{1}{2}A_{jk}^{(s)\dagger} A_{jk}^{(s)}\rho^{(s)}(t)-\frac{1}{2}\rho^{(s)}(t)A_{jk}^{(s)\dagger} A_{jk}^{(s)}\Big),
    \end{aligned}
\end{equation}
where $A_{jk}^{(s)}=|j\rangle\langle k|^{(s)}$ and $\gamma_{jk}^{(s)}$ are the scalar coefficients weighting each of the projectors $A_{jk}^{(s)}$. Meanwhile, the Hamiltonian $H^{(s)}(t)$ is given by
\begin{equation}
    H^{(s)}(t)=H_0^{(s)}+H_c^{(s)}(t),
    \label{specieshamiltonian}
\end{equation}
where $H_0^{(s)}$ describes the species time-independent drift Hamiltonian and $H_c^{(s)}(t)$ describes the species time-dependent control Hamiltonian. 

In the following, we consider a single species only and drop the superscript species label. The drift Hamiltonian for any given molecular species is given by,
\begin{equation}
\begin{aligned}
    H_0&=     T_{\text{nuc}}\big(\{\vec{{R}}_J\}\big)+ T_{\text{el}}\big(\{\vec{{r}}_j\}\big) + T_{\text{int}}\big(\{\vec{{R}}_J\},\{\vec{{r}}_j\}\big)\\
     &= \bigg\{\sum_{J=1}^{N_n}\frac{\vec{{P}}_J^2}{2m_J}+\sum_{J<K}\frac{Z_J Z_K}{\vec{R}_{JK}}\bigg\}+\bigg\{\sum_{j=1}^{N_e}\frac{\vec{{p}}_j^2}{2}+\sum_{j>k}\frac{1}{\vec{R}_{jk}}\bigg\}+\bigg\{-\sum_{J,j}\frac{Z_J}{\vec{R}_{Jj}}\bigg\},
\end{aligned}
\end{equation}
where $\{\vec{{R}}_J\}$ and $\{\vec{{r}}_j\}$ denote the sets of collective nuclear and electronic coordinates of a molecular species, respectively. Vector notation is used here to denote $\vec{{a}} = a_x\vec{{e}}_x+a_y\vec{{e}}_y+a_z\vec{{e}}_z$. We denote the charge and mass of the $J$-th nucleus by $Z_J$ and $m_J$ respectively, we denote the momentum of the $j$-th electron by $\vec{{p}}_{j}$ and the momentum of the $J$-th nucleus by $\vec{{P}}_{J}$, and we use the abbreviated notation $\vec{R}_{jk}\equiv |\vec{{r}}_j-\vec{{r}}_k|$, $\vec{R}_{JK}\equiv |\vec{{R}}_J-\vec{{R}}_K|$, and $\vec{R}_{Jj}\equiv |\vec{{R}}_J-\vec{{r}}_j|$ for the relative distances. The control Hamiltonian describes the interaction of the system with an applied laser field, and is modeled in the dipole approximation as
\begin{equation}
    H_c(t) = -\bigg(\sum_{j=1}^{N_e^{(s)}}\vec{{r}}_j^{(s)}-\sum_{J=1}^{N_n^{(s)}} Z_J^{(s)} \vec{{R}}_J^{(s)}\bigg)\cdot\vec{\varepsilon}(t),
\end{equation}
where $\vec{\varepsilon}(t)$ is the applied field amplitude at time $t$. The optical response of a single species $s$ is proportional to the acceleration of the dipole moment of the species, $\vec{R}(t) \propto \frac{d^2 }{dt^2}\big\langle \vec{\mu}\big\rangle_t$, where the notation $\langle \vec{O}\rangle_t\equiv {\rm Tr}\{O_x\rho(t)\}\vec{\mathbf{e}}_x +{\rm Tr}\{O_y\rho(t)\}\vec{\mathbf{e}}_y+{\rm Tr}\{O_z\rho(t)\}\vec{\mathbf{e}}_z$ denotes the expectation value of a vector operator at time $t$, and we have introduced the notation
\begin{equation}
    \vec{\mu}\equiv\sum_{j=1}^{N_e}\vec{{r}}_j-\sum_{J=1}^{N_n} Z_J\vec{{R}}_J
    \label{mu}
\end{equation}
for the dipole moment of the species. In the same spirit, we also adopt the following notation:
\begin{equation}
    \vec{p}\equiv \sum_{j=1}^{N_e}\vec{{p}}_j-\sum_{J=1}^{N_n} \frac{Z_J}{m_J}\vec{{P}}_J
    \label{p}
\end{equation}
\begin{equation}
        \vec{\nabla} \vec{V}\equiv \sum_{j=1}^{N_e}\vec{\nabla}_j\frac{Z_J}{\vec{R}_{Jj}}-\sum_{J=1}^{N_n}\frac{Z_J}{m_J}\vec{\nabla}_J\frac{Z_J}{\vec{R}_{Jj} }
        \label{nablaV}
\end{equation}
\begin{equation}
    Z_{tot}\equiv\sum_{J=1}^{N_n}Z_J
    \label{Ztot}
\end{equation}
\begin{equation}
    m_{tot}\equiv\sum_{J=1}^{N_n}m_J,
    \label{Ztot2}
\end{equation}
which leads to the following commutation relations:
\begin{equation}
\begin{aligned}
         [H(t),\vec{\mu} ]& =-i\vec{p},\qquad \qquad  [H(t),\vec{p}]&=i\vec{\nabla}\vec{ V}-i\big(N_e+\tfrac{Z_{tot}}{m_{tot}}\big)\vec{\varepsilon}(t),
    \label{commutationrelations}
    \end{aligned}
\end{equation}
where we remark that $\vec{\nabla}\vec{V} =  \frac{\partial V_x}{\partial x}\vec{\mathbf{e}}_x+\frac{\partial V_y}{\partial y}\vec{\mathbf{e}}_y+\frac{\partial V_z}{\partial z}\vec{\mathbf{e}}_z$. Our goal is to derive an expression for the species optical response when the system is acted on by the field $\vec{\varepsilon}(t)$. We first recall that from Eq. (\ref{DynamicalEq}), the time evolution of a species between times $t$ and $t+dt$ is given by
\begin{equation}
\rho(t+dt) = \mathcal{T} e^{\int_t^{t+dt} \mathcal{L} (\cdot) dt'}\rho(t),
\end{equation}
where $\mathcal{T}$ is the time-ordering operator. Then, using the Taylor expansion it can be shown that eliminating $\mathcal{T}$ contributes an error $O(dt^3)$. A Taylor expansion can also be used to show $\int_t^{t+dt}\mathcal{L}(\cdot) dt'= \mathcal{L}_D(\cdot) +O(dt^3) $ where $\mathcal{L}_D(\cdot)\equiv -i\Big[H\big(t+\tfrac{dt}{2}\big),\cdot\Big]dt+\mathcal{D}(\cdot)dt$, yielding
\begin{equation}
\langle \vec{\mu}\rangle_{t+dt} = {\rm Tr}\Big\{ \vec{\mu} e^{ \mathcal{L}_D (\cdot) dt}\rho(t)\Big\}+O(dt^3)\, .
\label{introduceLD}
\end{equation}
We now expand $e^{ \mathcal{L}_D (\cdot) dt}\rho(t)\approx \rho(t)+\mathcal{L}_D(\rho(t))dt+\mathcal{L}_D\big(\mathcal{L}_D(\rho(t))\big)\frac{dt^2}{2}+\cdots$. We truncate the expansion at second order, such that the error remains $O(dt^3)$. We can then use the definition of the adjoint operator
\begin{equation}
    {\rm Tr}\big\{\vec{\mu}\mathcal{L}_D(\rho(t))\big\}={\rm Tr}\big\{\mathcal{L}_D^\dagger(\vec{\mu})\rho(t)\big\},
\end{equation}
where 
\begin{equation}
    \mathcal{L}_D^\dagger(\vec{\mu})=i\Big[H\big(t+\tfrac{dt}{2}\big),\vec{\mu}\Big]+\mathcal{D}^\dagger(\vec{\mu})
    \label{Ldagger}
\end{equation}
and in the above, 
\begin{equation}
    \mathcal{D}^\dagger\big(\vec{\mu}\big)=\sum_{j,k}\gamma_{jk}\big(A_{jk}^\dagger\vec{\mu}A_{jk}-\frac{1}{2}A_{jk}^\dagger A_{jk}\vec{\mu}-\frac{1}{2}\vec{\mu}A_{jk}^\dagger A_{jk}\big),
\end{equation}
to write
\begin{equation}
    \begin{aligned}
\langle \vec{\mu}\rangle_{t+dt}&={\rm Tr}\big\{ \vec{\mu}\rho(t)\big\}+{\rm Tr}\bigg\{\Big(\mathcal{L}_D^\dagger( \vec{\mu})dt+\mathcal{L}_D^\dagger\big(\mathcal{L}_D^\dagger( \vec{\mu})\big)\tfrac{dt^2}{2}\Big)\rho(t)\bigg\}+O(dt^3)\\
&=\langle \vec{\mu}\rangle_t+\Big\langle\mathcal{L}_D^\dagger( \vec{\mu})dt+\mathcal{L}_D^\dagger\big(\mathcal{L}_D^\dagger( \vec{\mu})\big)\tfrac{dt^2}{2}\Big\rangle_t+O(dt^3).
\end{aligned}
\label{beforecommutator}
\end{equation}
We see that 
\begin{equation}
\begin{aligned}
    \mathcal{L}_D^\dagger\big(\mathcal{L}_D^\dagger( \vec{\mu})\big)    &=i\big[H\big(t+\tfrac{dt}{2}\big),\vec{p}\big]+\mathcal{D}^\dagger(\vec{p})+i\Big[H\big(t+\tfrac{dt}{2}\big),\mathcal{D}^\dagger(\vec{\mu})\Big]+\mathcal{D}^\dagger\big(\mathcal{D}^\dagger(\vec{\mu})\big).
    \label{LDLD}
\end{aligned}
\end{equation}
Substituting the relation from Eq. (\ref{commutationrelations}) into Eq. (\ref{LDLD}),
\begin{equation}
\begin{aligned}
    \mathcal{L}_D^\dagger\big(\mathcal{L}_D^\dagger( \vec{\mu})\big)&=-\vec{\nabla} \vec{V} +\big(N_e+\tfrac{Z_{tot}}{m_{tot}}\big)\vec{\varepsilon}(t+\tfrac{dt}{2})+\mathcal{D}^\dagger\big(\vec{p}+\mathcal{D}^\dagger(\vec{\mu})\big)+i\big[H\big(t+\tfrac{dt}{2}\big),\mathcal{D}^\dagger(\vec{\mu})\big].
    \label{LLdagger}
\end{aligned}
\end{equation}
From Eq. (\ref{Ldagger}), we see that $\frac{d\langle\vec{\mu}\rangle_t}{dt}=i\langle[H(t+\tfrac{dt}{2}),\vec{\mu}]\rangle_t+\langle\mathcal{D}^\dagger(\vec{\mu})\rangle_t=\langle \vec{p}\rangle_t+\langle\mathcal{D}^\dagger(\vec{\mu})\rangle_t$. We substitute this and the definition
\begin{equation}
    \langle \mathcal{\vec{A}}\rangle_t \equiv \Big\langle \mathcal{D}^\dagger\Big(\vec{p}+\mathcal{D}^\dagger(\vec{\mu})\Big)\Big\rangle_t+i\Big\langle \Big[H\big(t+\tfrac{dt}{2}\big),\mathcal{D}^\dagger(\vec{\mu})\Big]\Big\rangle_t\, .
\label{Ak} \end{equation}
into Eq. (\ref{LLdagger}). Then, substituting Eq. (\ref{LLdagger}) into Eq. (\ref{beforecommutator}) leads to
\begin{equation}
    \begin{aligned}
\langle \vec{\mu}\rangle_{(t+dt)}&= \langle \vec{\mu}\rangle_t+ dt \frac{d\langle\vec{\mu}\rangle_t}{dt} +\frac{dt^2}{2}\Big[\big(N_e+\tfrac{Z_{tot}}{m_{tot}}\big)\vec{\varepsilon}\big(t+\tfrac{dt}{2}\big) +\langle \mathcal{\vec{A}}\rangle_t -\langle \vec{\nabla}\vec{ V}\rangle_t \Big] +O(dt^3)\, ,
    \end{aligned}
\end{equation}
which can be rearranged to
\begin{equation}
    \begin{aligned}
\frac{2}{dt^2}\big(\langle \vec{\mu}\rangle_{(t+dt)}-\langle \vec{\mu}\rangle_t\big)&= \frac{2}{dt}\frac{d\langle\vec{\mu}\rangle_t}{dt}+    \big(N_e+\tfrac{Z_{tot}}{m_{tot}}\big)\vec{\varepsilon}\big(t+\tfrac{dt}{2}\big) +\langle \mathcal{\vec{A}}\rangle_t-\langle \vec{\nabla} \vec{V}\rangle_t +O(dt)\, .
    \end{aligned}
\end{equation}
We now substitute in the Taylor expansion $\langle \vec{\mu}\rangle_{(t+dt)}=\langle \vec{\mu}\rangle_t+ \frac{d \langle \vec{\mu}\rangle_t}{dt} dt+\frac{d^2 \langle \vec{\mu}\rangle_t}{dt^2} \frac{dt^2}{2}+O(dt^3)$
and cancel terms to arrive at our final expression for the species optical response $\vec{R}(t)=  \frac{d^2 \langle \vec{\mu}\rangle_t}{dt^2}$,
\begin{equation} 
\begin{aligned}
    \vec{R}(t)&=    \big(N_e+\tfrac{Z_{tot}}{m_{tot}}\big)\vec{\varepsilon}\big(t+\tfrac{dt}{2}\big) +\langle \mathcal{\vec{A}}\rangle_t-\langle \vec{\nabla} \vec{V}\rangle_t +O(dt).
\end{aligned} 
\label{Yj}
\end{equation}
We now return to our full system, which describes a mixture of $n_s$ species. The optical response of the mixture is defined as the sum of the optical responses of the species weighted by their relative concentrations, i.e.,
\begin{equation} 
\begin{aligned}
    \vec{R}_{\text{mix}}(t)&=\sum_{s=1}^{n_s}y_s\Big[     \big(N_e^{(s)}+\tfrac{Z_{tot}}{m_{tot}}^{(s)}\big)\vec{\varepsilon}\big(t+\tfrac{dt}{2}\big)+\langle \mathcal{\vec{A}}\rangle_t^{(s)}-\langle \vec{\nabla}^{(s)}\vec{V}^{(s)}\rangle_t  \Big] +O(dt).
    \label{Ymixunnormalized}
    \end{aligned}
\end{equation}
Molecular discrimination between species can be carried out by designing $n_s$ pulse shapes $\vec{\varepsilon}_1(t),\cdots,\vec{\varepsilon}_{n_s}(t)$ such that $\vec{\varepsilon}_s(t)$ suppresses the optical response of species $s$, i.e.,
\begin{equation}
\begin{aligned}
0
&=y_s\Big[     \big(N_e^{(s)}+\tfrac{Z_{tot}}{m_{tot}}^{(s)}\big)\vec{\varepsilon}_s\big(t+\tfrac{dt}{2}\big) +\langle \mathcal{\vec{A}}\rangle_t^{(s)}-\langle \vec{\nabla}^{(s)}\vec{V}^{(s)}\rangle_t    \Big],
\end{aligned}
\end{equation}
which rearranges to the final SSMC equation
\begin{equation} 
\vec{\varepsilon}_s(t+\tfrac{dt}{2})=-\frac{\Big( \langle \mathcal{\vec{A}}\rangle_t^{(s)}-\langle \vec{\nabla}^{(s)}\vec{V}^{(s)}\rangle_t \Big)}{ \big(N_e^{(s)}+\tfrac{Z_{tot}}{m_{tot}}^{(s)}\big) },
\label{field}
\end{equation}
which is free of singularities due to the fact that the denominator is restricted to be a positive number. For a closed quantum system undergoing unitary dynamics, the expression would be the identical, but with $\langle \mathcal{\vec{A}}\rangle_t^{(s)} = 0$.\footnote{The molecular case discussed in the main text in Sec. \ref{sec:molcharacterisation} can be obtained by assuming closed system dynamics, and by assuming the Born-Oppenheimer approximation, such that the dynamics of the nuclei and electrons are assumed to be uncoupled. Then, the illustrations in Sec. \ref{Sec:molecularnumerics} use a model for the vibrational dynamics of the nuclei alone.}

\end{document}